\documentclass[twocolumn,english,pra,showpacs,superscriptaddress,longbibliography]{revtex4-1}
\usepackage[english]{babel}  
\usepackage{amsfonts}
\usepackage{amsmath}
\usepackage{graphics}
\usepackage{amssymb}
\usepackage{physics}
\usepackage[colorlinks, citecolor=blue,linkcolor=red]{hyperref}
\usepackage{color}
\usepackage{hyperref}
\usepackage{textcomp}
\usepackage[rightcaption]{sidecap}
\usepackage{subfigure}
\usepackage{float,csquotes,mathtools}
\usepackage[rightcaption]{sidecap}
\usepackage{graphicx}
\usepackage{lipsum}

\begin{document}
\title{Coherent Control of the Goos–H\"{a}nchen Shift in Polariton Optomechanics}
\author{Shah Fahad}
\affiliation{Department of Physics, Zhejiang Normal University, Jinhua, Zhejiang 321004, China}
\author{Gao Xianlong}
\email {gaoxl@zjnu.edu.cn}\affiliation{Department of Physics, Zhejiang Normal University, Jinhua, Zhejiang 321004, China}

\setlength{\parskip}{0pt}
\setlength{\belowcaptionskip}{-10pt}
\begin{abstract}
We propose a theoretical scheme for controlling the Goos–H\"{a}nchen shift (GHS) of a reflected probe field in a polariton optomechanical system. The system comprises an optical mode, a molecular vibrational mode, and $N$ excitonic modes, where excitons couple to molecular vibrations via conditional displacement interactions and to photons through electric dipole interactions. We show that the effective exciton–vibration coupling provides a powerful mechanism for coherent GHS control: in its absence, the system exhibits a pronounced GHS at resonance, while activating it strongly suppresses the shift. The effective cavity detuning and the cavity length serve as additional tunable parameters for GHS manipulation. Furthermore, increasing the collective exciton–optical coupling enhances the GHS. Our results establish a framework for probing the GHS in polariton optomechanical systems and offer new avenues for designing optical devices that exploit beam-displacement phenomena.
\end{abstract}
\date{\today}
\maketitle
\section{Introduction}
Quantum cavity optomechanics has emerged as a rapidly developing branch of modern physics with broad interdisciplinary significance~\cite{Aspelmeyer-book-2014, Aspelmeyer2014}. This field applies the principles and techniques of quantum optics to investigate light–matter interactions across a wide range of condensed-matter platforms. In particular, dispersively coupled optomechanical systems provide an effective platform for realizing quantum memories and quantum information carriers~\cite{Aspelmeyer2014, Kippenberg, Metcalfe-2014}. Based on these developments, recent advances in quantum technologies~\cite{Barzanjeh2022}, quantum control~\cite{Connell2010,Martin-2017, Yiwen-2017}, quantum entanglement~\cite{Mancini-2002, Vitali-2007, Paternostro-2007}, ground-state cooling~\cite{Wilson-PRL, Marquardt-PRL,Jasper-2011}, nanoscale optical cavities~\cite{Verhagen2012}, and optomechanically induced transparency (OMIT)~\cite{Stefan-2010} have further expanded the capabilities of cavity optomechanical systems. OMIT is formally analogous to electromagnetically induced transparency (EIT) in a cavity optomechanical system operating within the resolved sideband regime~\cite{Schliesser-2009, Schliesser-2010, Agarwal-2010}.

Molecular cavity optomechanics, which investigates the coupling between electromagnetic fields and molecular vibrational modes, has recently emerged as a promising platform for realizing room-temperature quantum control~\cite{Roelli2016, Schmidt2016, Hughes2021, Esteban2022}. Compared to conventional optomechanical systems, it offers two major advantages: (i) the inherently high-frequency vibrational modes of molecules (on the order of tens of terahertz) allow them to remain in their quantum ground state even at elevated temperatures, and (ii) embedding molecules within the ultrasmall mode volume of a plasmonic cavity enables strong photon–vibration coupling. However, the strong optical confinement in plasmonic cavities induces significant dissipation $(\kappa \sim 33~\mathrm{THz})$~\cite{Roelli2016}, requiring high pump powers to maintain strong optomechanical coupling—conditions that may damage molecular vibrational bonds~\cite{Anna-PRX}. To address these limitations, a molecular quantum optomechanical platform coherently integrates excitonic, photonic, and vibrational modes~\cite{Shishkov-Room, Shishkov-PRL}. In this setup, the excitonic mode exhibits strong coupling to both a high-Q optical cavity mode and a molecular vibrational mode, giving rise to a tripartite polariton–optomechanical interaction~\cite{Shishkov-PRL, Shishkov-Room, Carlon-PRL, Jusserand-PRL, Kyriienko-PRL, Jian-2025}. In this work, we investigate the Goos–H\"{a}nchen shift (GHS) as a distinct and measurable signature of the optical response in a polariton optomechanical system, revealing how phase-sensitive beam shifts can serve as probes of the system’s internal dynamics. 

When a light beam undergoes total internal reflection at the interface between two dissimilar media, it experiences a lateral displacement known as GHS~\cite{Picht-1929, Goos-1943, Goos-1947}. The GHS has garnered attention due to its relevance in high-precision optical sensing~\cite{Yi-2008, Hashimoto:89}, including temperature-dependent sensors~\cite{Chen-07}, and plasmonic sensors~\cite{Yin-2006}. Both positive and negative GHS have been observed under diverse dispersive conditions. Positive shifts typically occur in conventional total internal reflection~\cite{Renard:64}, at interfaces between left- and right-handed media~\cite{Qing:04}, and within periodic photonic structures~\cite{Tamir:71}, whereas negative shifts are often associated with absorbing materials~\cite{Chu-82} or media exhibiting negative refractive indices~\cite{Birman-83, Berman-02}.  

The enhancement of the GHS has been extensively investigated across a wide range of optical platforms, including absorptive media~\cite{Pfleghaar-93}, photonic crystals~\cite{Soboleva-2012, Wan:12}, structural resonances~\cite{Yang:14}, metallic interfaces~\cite{Bonnet:01, Yin-2004}, and engineered metamaterials~\cite{Xu-2015, Chen_2016}. More recently, research on the GHS has expanded into quantum and hybrid light–matter systems, including quantum dots~\cite{Idress-PRA}, atom–cavity quantum electrodynamics setups~\cite{Zia_2010_Coherent}, cavity optomechanical platforms~\cite{Muhib-PRA, Anwar-PRA}, and cavity magnomechanics~\cite{waseem_Goos_2024, fahad2025}.

In this paper, we investigate the manipulation of the GHS of a reflected probe field in a polariton optomechanical system using the stationary-phase method. The hybrid system comprises $N$ excitonic modes, an optical mode, and a molecular vibrational mode. The optical mode couples to excitons via electric dipole interactions, while excitons interact with molecular vibrations through conditional displacement interactions. We demonstrate that the GHS can be coherently controlled by varying the effective exciton–vibration coupling strength. In the absence of this effective coupling, a pronounced GHS appears at resonance, whereas introducing the effective coupling significantly suppresses the GHS. Moreover, we show that the GHS can be precisely controlled by adjusting the effective cavity detuning. Notably, in the absence of effective exciton–vibration coupling, adjusting the detuning allows a smooth transition of the GHS from positive to negative values. In contrast, when the effective exciton–vibration coupling is present, the GHS extends over a wide range of negative detuning values. We further demonstrate that tuning the intracavity length provides an additional degree of control over the GHS. Finally, we show that increasing the collective exciton–optical coupling enhances the GHS for a fixed effective exciton–vibration coupling strength.

This paper is organized as follows. Section II develops the theoretical model. Starting from the system Hamiltonian, we derive the linearized Heisenberg–Langevin equations, obtain the optical susceptibility, and finally employ the stationary phase method to compute the GHS. Section III presents the numerical results and discussion, and Section IV concludes the paper.

\section{System Model and Hamiltonian}
\subsection{Polariton optomechanics
and Heisenberg–Langevin equations}
\begin{figure}[tp]
\includegraphics[width=\linewidth]{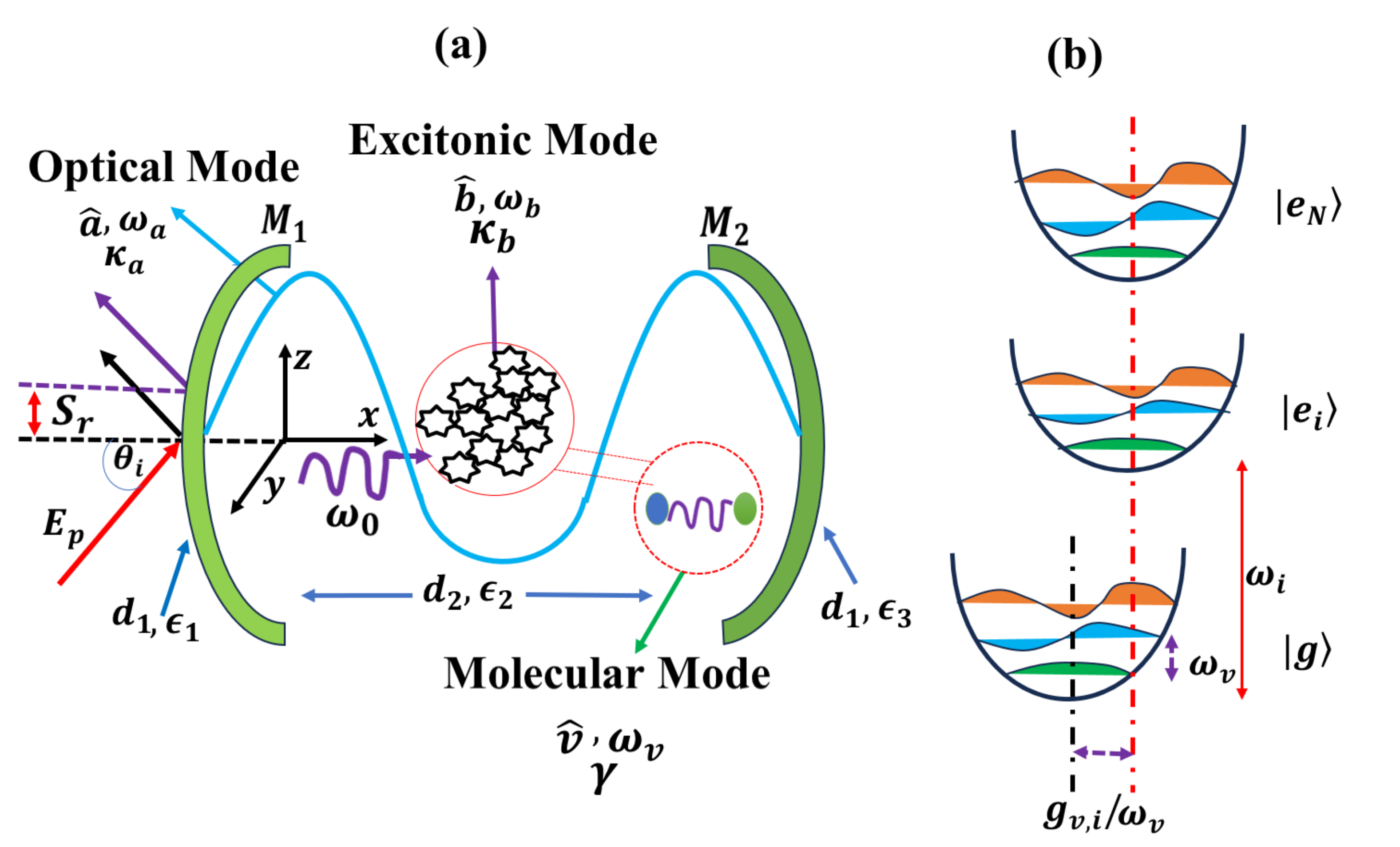}
\caption{Schematic illustration of the polariton optomechanical system. (a) A single-mode optical cavity (frequency $\omega_{a}$, decay rate $\kappa_{a}$) interacts with $N$ excitonic modes (frequency $\omega_{b}$, decay rate $\kappa_{b}$) through electric-dipole coupling. The excitonic modes further couple to the vibrational mode of the molecule (frequency $\omega_v$, decay rate $\gamma$) via conditional displacement interactions. A pump field (frequency $\omega_{0}$, amplitude $\eta$) drives the excitonic modes. A transverse electric (TE)-polarized probe field $E_{p}$ incident on mirror $M_{1}$ at an angle $\theta_{i}$ acquires a lateral displacement upon reflection, referred to as the GHS, and represented by $S_{r}$. (b) Simplified energy level diagram illustrating the conditional displacement interaction between the electronic transition and the vibrational state.}
\label{fig1}
\end{figure}
We present a polariton optomechanical platform in which a single-mode optical cavity is coherently coupled to a large organic molecule featuring vibrationally dressed electronic transitions, as illustrated in Fig.~\ref{fig1}(a) and (b).  The molecule hosts $N$ electronically excited states $\lvert e_{i} \rangle$, each connected to a common ground state $\lvert g \rangle$ through optical transitions~\cite{Reitz-2020, Mart-PRL}. These electronic transitions are linearly coupled to a molecular vibrational mode via conditional displacement coupling and simultaneously to the cavity optical mode through electric-dipole coupling. This setup facilitates an effective polariton optomechanical coupling among three distinct resonant modes~\cite{Shishkov-PRL, Shishkov-Room}. The optical cavity is formed by two nonmagnetic mirrors, $M_{1}$ and $M_{2}$, each with a thickness $d_{1}$. $M_{1}$ (permittivity $\epsilon_{1}$) is partially reflective, whereas $M_{2}$ (permittivity $\epsilon_{3}$) is perfectly reflective. The mirrors are separated by a distance $d_{2}$, with an intracavity medium of effective permittivity $\epsilon_{2}$ between them.

The system Hamiltonian ($\hbar=1$) is expressed as~\cite{Jian-2025}
\begin{align}
\label{Main-H}
\hat{H}& = 
\sum_{i=1}^N \omega_{i}\hat{\xi}_{i}^\dagger \hat{\xi}_{i}
+ \omega_{a} \hat{a}^\dagger \hat{a}
+ \omega_{v} \hat{v}^\dagger \hat{v}\\
&\quad+ \sum_{i=1}^N g_{v,i} \hat{\xi}_{i}^\dagger \hat{\xi}_{i} \left(\hat{v}^\dagger + \hat{v}\right)
+ \sum_{i=1}^N g_{a,i}\left(\hat{\xi}_{i}^\dagger \hat{a} + \hat{\xi}_{i} \hat{a}^\dagger\right)\nonumber\\
&\quad+ \eta \sum_{i=1}^N \left(\hat{\xi}_i^\dagger e^{-i\omega_{0} t} + \hat{\xi}_i e^{i\omega_{0} t}\right)
+ iE_{p} \left(\hat{a}^\dagger e^{-i\omega_{p} t} - \hat{a} e^{i\omega_p t}\right),\nonumber
\end{align}
where $\hat{\xi}_{i} = \ket{g}\bra{e_i}$ ($\hat{\xi}_{i}^\dagger$) denotes the annihilation (creation) operator of the $i$th electronic transition with resonance frequency $\omega_i$. Unlike a standard two-level system, these $N$ transitions arise from a single ground state. This system can be modeled as an optical mode (resonance frequency $\omega_a$, decay rate $\kappa_a$) coupled to a molecular mode with a single vibrational mode (resonance frequency $\omega_v$, decay rate $\gamma$) and $N$ distinct electronic states, each treated as an effective two-level system, leading to a Hamiltonian of the form in Eq.~(\ref{Main-H}). The optical and molecular vibrational modes are described by the annihilation (creation) operators $\hat{a}$ ($\hat{a}^\dagger$) and $\hat{v}$ ($\hat{v}^\dagger$), respectively, which obey the canonical commutation relations
$[\hat{a}, \hat{a}^\dagger] = [\hat{v}, \hat{v}^\dagger] = 1$ and $[\hat{\xi}_{i}, \hat{\xi}_{j}^\dagger] = \delta_{ij}\ket{g}\bra{g} - \ket{e_j}\bra{e_i}$. $g_{v,i}$ characterizes the coupling between the $i$th electronic transition and the molecular vibrational mode, and is related to the square root of the Huang-Rhys parameter~\cite{Spano2010}, whereas $g_{a,i}$ represents the coupling between the optical mode and the electronic transitions. The external pump field (frequency $\omega_0$ and amplitude $\eta$) drives the electronic transitions. Furthermore, the optical cavity is probed by a weak field (frequency $\omega_{p}$, amplitude $E_{p}$). Using the unitary transformation $\hat{\tilde{H}} = \hat{U}^\dagger \hat{H} \hat{U} - i\hat{U}^\dagger \frac{\partial \hat{U}}{\partial t}$, with $\hat{U} = \exp[-i\omega_{0} (\hat{a}^\dagger \hat{a} + \sum_{i=1}^N \hat{\xi}_{i}^\dagger \hat{\xi}_{i}) t]$, the system Hamiltonian is written as:
\begin{align}
\hat{\tilde{H}} 
&= \sum_{i=1}^N \Delta_{i}\hat{\xi}_i^\dagger \hat{\xi}_{i} + \Delta_{a}\hat{a}^\dagger \hat{a} + \omega_{v}\hat{v}^\dagger \hat{v} + \sum_{i=1}^N g_{v,i}\hat{\xi}_{i}^\dagger \hat{\xi}_{i}(\hat{v}^\dagger + \hat{v}) \nonumber\\
&\quad + \sum_{i=1}^N g_{a,i}\left(\hat{\xi}_{i}^\dagger \hat{a} + \hat{\xi}_{i} \hat{a}^\dagger\right)+ \eta\sum_{i=1}^N\left(\hat{\xi}_{i}^\dagger + \hat{\xi}_{i}\right) \nonumber\\
&\quad + iE_{p}\left(\hat{a}^\dagger e^{-i\delta_{p} t} - \hat{a} e^{i\delta_{p} t}\right),
\label{H-RTW}
\end{align}
where $\Delta_{i} = \omega_{i} - \omega_{0}, \quad \Delta_{a} = \omega_{a} - \omega_{0}$, and $\delta_{p} =\omega_{p}-\omega_{0}$ are the detunings of the $i$th electronic transition, the optical mode, and the probe field relative to the driving field. To simplify the analysis, we assume uniform coupling strengths and transition frequencies across all electronic states, such that $\omega_{i} \approx \omega_b$, $g_{v, i} \approx g_v$, and $g_{a, i} \approx g_{a}$. Under low excitation and large detuning, the electronic transitions are well-described by collective low-energy excitations. We therefore define a collective bright mode $\hat{b} = \sum_{i=1}^N \hat{\xi}_{i}/\sqrt{N}$ for the $N$ excitations, which obeys $[\hat{b}, \hat{b}^\dagger] = 1$~\cite{Kurucz-PRA}. The $N-1$ dark modes are obtained recursively as follows:
$\hat{b}_{1} = \hat{\xi}_{1}, \quad
\hat{b}_{2} = ({\hat{\xi}_{1} + \hat{\xi}_2})/{\sqrt{2}}, \quad
\hat{b}_{i-1} = (\sqrt{i-2} \hat{b}_{i-2} + \hat{\xi}_{i-1})/{\sqrt{i-1}}$ $(i=3,\dots,N)$. The dark modes are then defined as
$\hat{d}_{i} = (\hat{b}_{i-1} - \sqrt{i-1} \hat{\xi}_{i})/{\sqrt{i}}$ $(i=2,\dots,N)$~\cite{huang2023}. The total number of excitations can thus be expressed as $\sum_{i=1}^{N} \hat{\xi}_i^\dagger \hat{\xi}_{i}= \hat{b}^\dagger \hat{b} + \sum_{i=2}^{N} \hat{d}_i^\dagger \hat{d}_{i}$. Under the condition of large off-resonant driving~\cite{Jian-2025, Mart-PRL}, the dark modes remain unpopulated and can be neglected, reducing the description to the single bright collective mode. Equation~(\ref{H-RTW}) can then be written as:
\begin{align}
\hat{\tilde{H}} &\simeq 
\Delta_{b} \hat{b}^\dagger \hat{b}
+ \Delta_{a} \hat{a}^\dagger \hat{a}
+ \omega_{v}\hat{v}^\dagger \hat{v} 
+ g_{v} \hat{b}^\dagger \hat{b}(\hat{v}^\dagger + \hat{v})\\
&\quad+ G_{a} (\hat{b}^\dagger \hat{a} + \hat{b}\hat{a}^\dagger)
+ \Omega(\hat{b}^\dagger + \hat{b})
+ iE_{p}\left(\hat{a}^\dagger e^{-i\delta_{p} t} - \hat{a} e^{i\delta_{p} t}\right),\nonumber
\label{simplified-H}
\end{align}
where $\Delta_{b} = \omega_{b} - \omega_{0}$, $G_a = \sqrt{N}g_{a}$ represents the collective exciton–optical mode coupling strength, and $\Omega = \sqrt{N}\eta$ is the effective drive amplitude.

We derive the Heisenberg-Langevin equations of motion $\frac{d\hat{O}}{dt} = i [\hat{\tilde{H}}, \hat{O}] - \Gamma \hat{O}$, where $\Gamma$ is the decay ($\Gamma>0$) or gain ($\Gamma<0$) rate for an operator $\hat{O}$ satisfying $[\hat{O}, \hat{O}^\dagger] = 1$. To analyze the average cavity response, we neglect quantum and thermal noise and describe the system dynamics for $\hat{O} \in \{\hat{a}, \hat{b}, \hat{v}\}$ as:
\begin{align}
\dot{\hat{a}}&= -(i\Delta_{a}+\kappa_{a})\hat{a}-iG_{a}\hat{b} + E_{p}e^{-i\delta_{p} t},\nonumber\\    
\dot{\hat{b}}&= -(i\Delta_{b} + \kappa_{b})\hat{b} -ig_{v}\hat{b}(\hat{v}^\dagger+\hat{v})-iG_{a}\hat{a} -i \Omega,\nonumber\\ 
\dot{\hat{v}}&= -(i\omega_{v}+\gamma)\hat{v} -ig_{v}\hat{b}^{\dagger}\hat{b},
\end{align}
where $\kappa_{a}$, $\kappa_{b}$, and $\gamma$ correspond to the dissipation rates of the optical, excitonic, and molecular vibrational modes, respectively. Replacing the operators with their expectation values $O(t) \equiv \langle \hat{O}(t) \rangle$ ($O = a, b, v$)~\cite{Xiong2015} yields:
\begin{align}
\dot{a}&= -(i\Delta_{a}+\kappa_{a})a-iG_{a}b + E_{p}e^{-i\delta_{p} t},\nonumber\\    
\dot{b}&= -(i\Delta_{b} + \kappa_{b})b -ig_{v}b(v^{\ast} +v)-iG_{a}a -i \Omega,\nonumber\\ 
\dot{v}&= -(i\omega_{v}+\gamma)v -ig_{v}b^{\ast}b. 
\end{align}
For a probe field significantly weaker than the pump, each operator’s expectation value is written as $O = O_{s} + \delta O$, with $O \in \{a, b, v\}$. The steady-state values $O_{s}$ are:
\begin{align}
a_{s}&= \frac{-iG_{a}b_{s}}{( i\Delta_{a}+\kappa_{a})},\nonumber\\
b_{s}& = \frac{-iG_{a}{a}_{s} -i\Omega}{(i\tilde\Delta_{b} + \kappa_{b})},\nonumber\\
v_{s} &= \frac{-ig_{v} |{b_{s}}|^2}{(i\omega_{v}+\gamma)},\label{steady states}
\end{align}
where $\tilde\Delta_{b}= \Delta_{b} + g_{v}({v}^{\ast}_{s} +{v_s})$ is the effective excitation drive detuning. The first-order linearized fluctuation terms $\delta O$ are:
 \begin{align}
\delta\dot{a}& = -(i\Delta_{a} + \kappa_{a})\delta{a}-iG_{a}\delta{b} + E_{p}e^{-i \delta_{p} t},\nonumber\\
\delta\dot{b}&= - (i\tilde\Delta_{b} + \kappa_{b})\delta{b}-iG_{v} \delta{v}-iG_{a}\delta{a},\nonumber\\ 
\delta\dot{v}& =- (i\omega_{v} + \gamma) \delta{v}-iG_{v}^\ast\delta{b}. \label{linerized-1}  
 \end{align}
Here, $G_{v}=g_{v}b_{s}$ is the effective exciton–vibration coupling strength. Since the control field is tuned to the red sideband, the terms $iG_{v}\delta v^\ast$ and $iG_{v}\delta b^\ast$ are off-resonant and can be omitted under the rotating wave approximation.

\subsection{Optical Susceptibility}
To investigate the reflection properties of the weak probe field, we compute the optical susceptibility $\chi$ of the intracavity of the polariton optomechanical system. We solve the linearized equation of motion Eq.~(\ref{linerized-1}) by transforming to a frame rotating with the slowly varying amplitudes of the fluctuations. This transformation is implemented through the substitutions $\delta a \rightarrow \delta a e^{-i\Delta_{a}t}$, $\delta b \rightarrow \delta b e^{-i\tilde\Delta_{b}t}$, and $\delta v \rightarrow \delta v e^{-i\omega_{v}t}$. Furthermore, the system operates under the red-sideband condition, where the pump field driving the excitons is detuned such that $\Delta_{a} = \tilde\Delta_{b} = \omega_{v}$~\cite{Agarwal-2010}. Under these conditions, Eq.~(\ref{linerized-1}) takes the form:
\begin{equation}
 \begin{aligned}
\delta\dot{a}& = -\kappa_{a}\delta a-iG_{a}\delta b + E_{p}e^{-ixt},\\
\delta\dot{b}&= -\kappa_{b}\delta {b}-iG_{v}\delta{v} -iG_{a}\delta a,\\ 
\delta\dot{v}& = - \gamma \delta{v} -iG_{v}^{\ast}\delta b, \label{linerized-2} 
 \end{aligned}
\end{equation}
where $x=\delta_{p}-\omega_{v}$ is the effective detuning. To solve Eq.~(\ref{linerized-2}), we apply the ansatz $\delta O = \delta{O}_{1}e^{-ixt} + \delta O_{2}e^{ixt}$ with $O= (a, b,v)$ and derive the following expression for the amplitude $\delta a_{1}$ of the first-order sideband of the weak probe field:
\begin{equation}
\delta{a}_{1} = \frac{E_{p}}{(\kappa_{a}-ix) + \frac{(\gamma-ix)G_{a}^2}{(\kappa_{b}-ix)(\gamma -ix) + |G_{v}|^2}}.\label{amplitude} 
\end{equation}
The contribution of $\delta a_{2}$ is negligible, as it stems from a four-wave mixing process at frequency $\omega_{p} - 2\omega_{0}$, mediated by the interaction between the weak probe and the driving fields. The output field $E_{T}$ of the weak probe defines the optical susceptibility $\chi$ via the relation~\cite{waseem_Goos_2024, Chen_2023,Li_2016Transparency, fahad2025}
\begin{equation}
\chi \equiv E_T = \kappa_{a} \delta a_{1} / E_{p},\label{optical sus}
\end{equation}
where the $\chi = \chi_{r} + i\chi_{i}$ represents the complex susceptibility, with its quadrature components measurable via homodyne detection~\cite{walls1994quantum}. The real part $\chi_{r}$ corresponds to the absorption spectrum of the weak probe field, while the imaginary part $\chi_{i}$ describes its dispersion spectrum. The effective intracavity permittivity $\epsilon_{2}$ reflects the cavity’s response to the probe field and is related to the optical susceptibility via $\epsilon_{2} = 1 + \chi$.
\subsection{Goos-H\"{a}nchen shift}
When the transverse electric (TE)-polarized probe field $E_{p}$ reflects from the mirror $M_{1}$, it experiences a lateral shift along the $z$-axis, referred to as the GHS $S_{r}$. To quantify this shift, we adopt the stationary phase method, which treats a highly collimated probe beam with a narrow angular spectrum as an approximate plane wave. Under this formalism, the lateral displacement of the reflected field can be expressed as~\cite{Artmann-1948, Li-2003PRL}
\begin{equation}
S_{r} = -\frac{\lambda} {2\pi} \frac{d\phi_{r}}{d\theta_{i}}.
\end{equation}
Here, $\lambda$ denotes the probe field wavelength, $\phi_{r}$ corresponds to the phase of the TE-polarized reflection coefficient $r(k_{z}, \omega_{p})$, $k_{z} = (2\pi / \lambda) \sin{\theta_{i}}$ represents the $z$-component of the wavenumber, and $\theta_{i}$ is the angle of incidence. The GHS can be expressed explicitly as~\cite{Wang:05}
\begin{equation}
\begin{split}
S_{r} = -\frac{\lambda}{2\pi} \frac{1}{|r(k_{z},\omega_{p})|^2}
\left\{\operatorname{Re}[r(k_{z},\omega_{p})] \frac{d \operatorname{Im}[r(k_{z}, \omega_{p})]}{d\theta_{i}}\right. \\
\left. - \operatorname{Im}[r(k_{z}, \omega_{p})] \frac{d \operatorname{Re}[r(k_{z}, \omega_{p})]}{d\theta_{i}}
\right\}.
\label{GHS} 
\end{split}
\end{equation}
The reflection coefficient $r(k_{z}, \omega_{p})$ in Eq.~(\ref{GHS}) is obtained using the standard transfer-matrix method~\cite{Wang_2008} and reads 
\begin{equation}
r(k_{z}, \omega_{p}) = \frac{q_{0}(Q_{22} - Q_{11}) - (q_{0}^2 Q_{12} - Q_{21})}{q_{0}(Q_{22} + Q_{11}) - (q_{0}^2 Q_{12} + Q_{21})}, \label{2x2}
\end{equation}
where $q_{0} = \sqrt{\epsilon_{0} - \sin^2{\theta_{i}}}$, and $Q_{ij}$ ($i,j = 1,2$) denote the elements of the total transfer matrix $Q(k_{z}, \omega_{p})$. The total transfer matrix of the effective three-layer polariton optomechanical system is then given by~\cite{Wang_2008}
\begin{widetext}
\begin{equation}   
Q(k_{z}, \omega_{p}) = M_{1}(k_{z}, \omega_{p}, d_{1}) M_{2}(k_{z}, \omega_{p}, d_{2}) M_{1}(k_{z}, \omega_{p}, d_{1}) =\begin{pmatrix}
Q_{11} & Q_{12} \\
Q_{21} & Q_{22}
\end{pmatrix}.
\end{equation}
\end{widetext}
Here, $M_{j}(k_{z}, \omega_{p}, d_{j})$ is the transfer matrix associated with the $j$-th layer
\begin{equation}
M_{j}(k_{z},\omega_{p},d_{j}) =
\begin{pmatrix}
\cos[k_{x}^{j} d_j] & ik/ k_{x}^{j} \sin[k_{x}^{j} d_{j}] \\
i k_{x}^{j}/k \sin[k_{x}^{j} d_{j}] & \cos[k_{x}^{j} d_{j}]
\end{pmatrix}, \label{transfer matrix}
\end{equation}
where $k_{x}^{j} = k \sqrt{\epsilon_{j} - \sin^2{\theta_{i}}}$ is the $x$-component of the probe field wavenumber in the $j$-th layer. Here,
$k = \omega_{p}/c$ is the vacuum wave number and and $c$ is the speed of light. Each layer ($j\equiv
1,2$) is characterized by its permittivity $\epsilon_{j}$ and thickness $d_{j}$. This transfer-matrix approach is widely adopted in GHS studies, with applications in atomic systems~\cite{Wang_2008, Zia_2010_Coherent}, cavity optomechanics~\cite{Ghaisuddin-2021}, and cavity magnomechanics~\cite{waseem_Goos_2024}.
\section{Results and Discussion}
\begin{figure}
\begin{center}
\includegraphics[width=\linewidth]{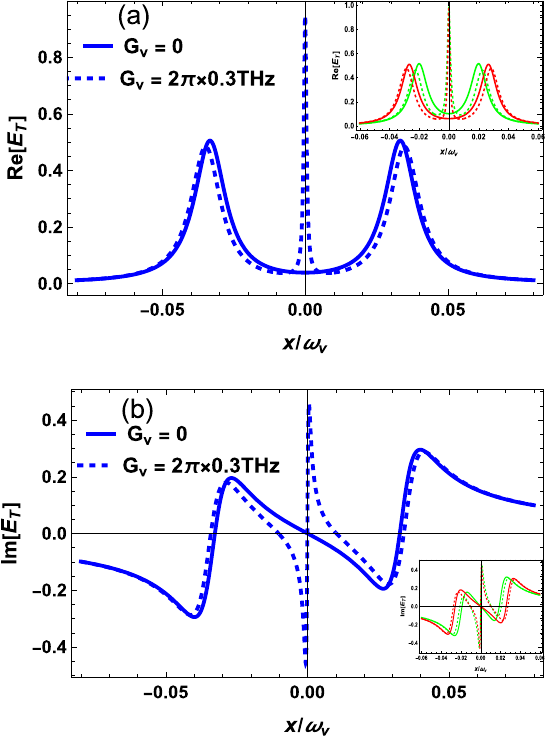}
\caption{(a) Absorption spectra $\mathrm{Re}[E_T]$ and (b) dispersion spectra $\mathrm{Im}[E_T]$ of the probe field versus the normalized effective detuning $x/\omega_v$. The solid blue curves represent the absence of effective exciton-vibration coupling ($G_v=0$), while the dashed blue curves correspond to $G_v=2\pi\times0.3~\mathrm{THz}$ at a fixed collective exciton–optical coupling strength $G_{a}/2\pi = 1.0~\mathrm{THz}$. The insets in Figs.~\ref{fig2}(a) and (b) show the transparency-window width for $G_{a}/2\pi=0.6~\mathrm{THz}$ (green) and $G_{a}/2\pi=0.8~\mathrm{THz}$ (red). Fixed parameters: $\omega_v/2\pi = 30~\mathrm{THz}$, $\kappa_{a}/2\pi = \kappa_{b}/2\pi = 0.2~\mathrm{THz}$, and $\gamma/2\pi = 1.0~\mathrm{GHz}$.} \label{fig2}
\end{center}
\end{figure}
\begin{figure}
\begin{center}
\includegraphics[width=\linewidth]{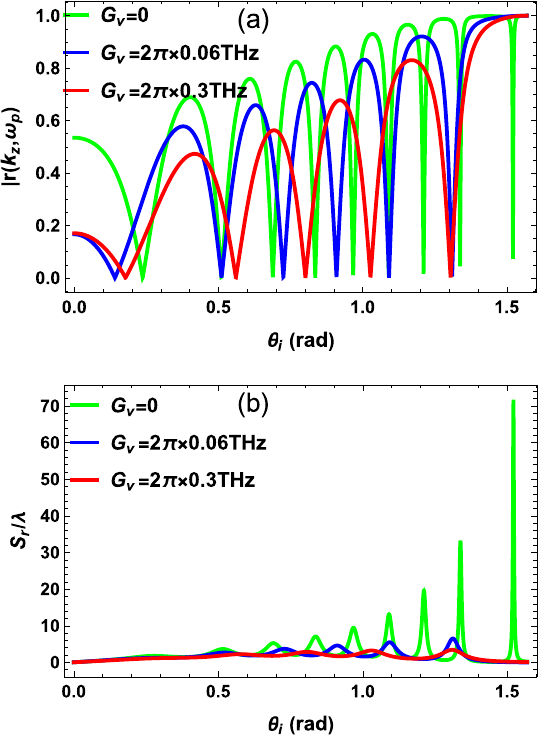}
\caption{(a) Absolute value of the reflection coefficient $|r(k_{z}, \omega_{p})|$ and (b) the normalized GHS $S_{r}/\lambda$ versus incident angle of the probe field $\theta_{i}$, for different effective exciton–vibration coupling strengths: $G_{v} = 0$ (green), $2\pi\times0.06~\mathrm{THz}$ (blue), and $2\pi\times0.3~\mathrm{THz}$ (red), respectively, at the resonance condition ($x=0$). Fixed parameters: $\omega_a/2\pi = 300~\mathrm{THz}$,  $\omega_v/2\pi = 30~\mathrm{THz}$, $G_{a}/2\pi = 1.0~\mathrm{THz}$, $\kappa_{a}/2\pi = \kappa_{b}/2\pi = 0.2~\mathrm{THz}$, $\gamma/2\pi = 1.0~\mathrm{GHz}$, $\epsilon_0=1$, $\epsilon_1=\epsilon_3=2.22$, $d_1=0.2~\mu\mathrm{m}$, and $d_2=5~\mu \mathrm{m}$.}
\label{fig3}
\end{center}
\end{figure}
\begin{figure}
\centering
\includegraphics[width=\linewidth]{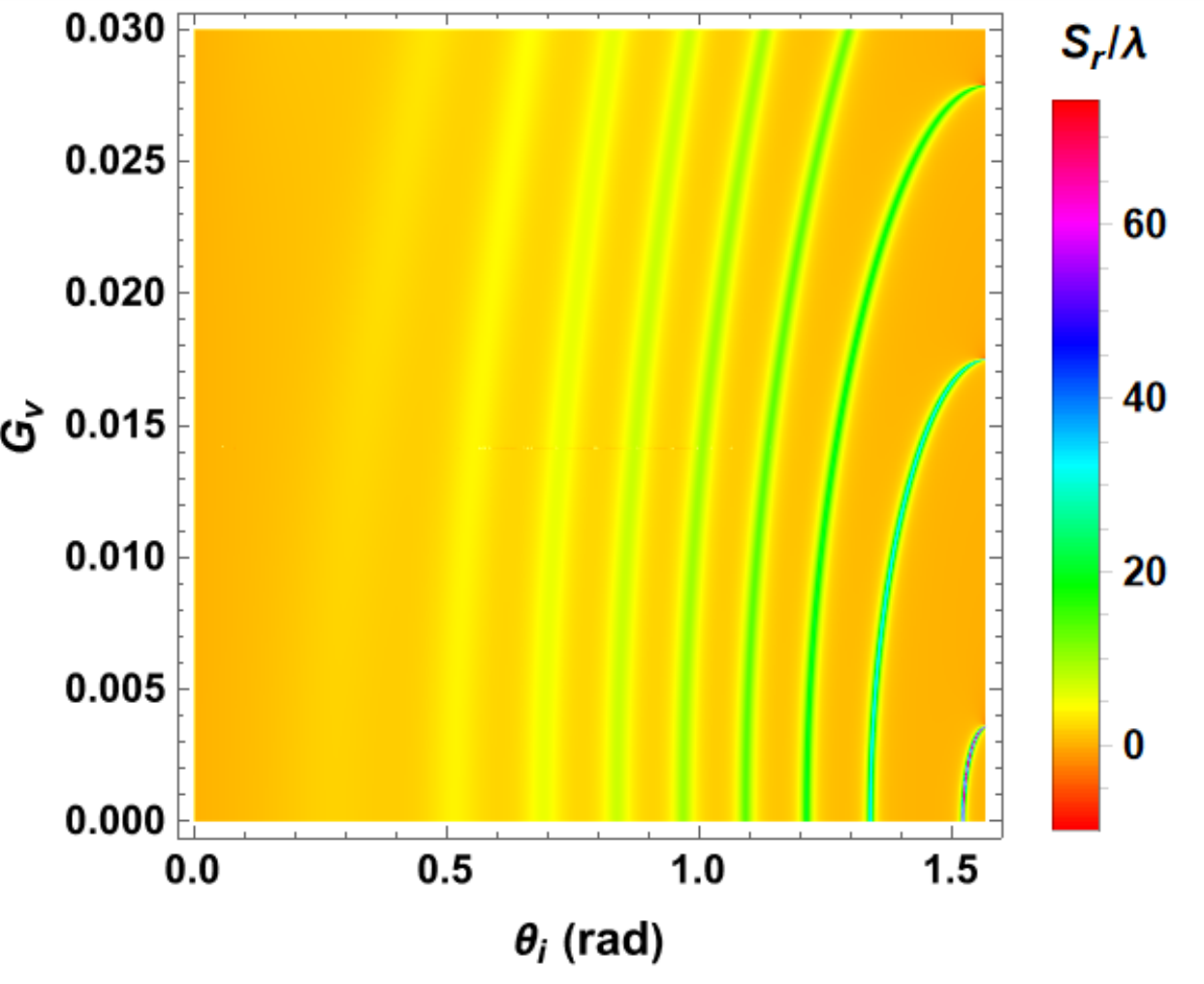}
\caption{Normalized GHS $S_{r}/\lambda$ as functions of incident angle $\theta_{i}$ and effective exciton–vibration coupling strength $G_{v}$ (units: $2\pi\times1.0~\mathrm{THz}$) at resonance ($x=0$).  Fixed parameters: $\omega_a/2\pi = 300~\mathrm{THz}$, $\omega_v/2\pi = 30~\mathrm{THz}$, $G_{a}/2\pi = 1.0~\mathrm{THz}$, $\kappa_{a}/2\pi = \kappa_{b}/2\pi = 0.2~\mathrm{THz}$, $\gamma/2\pi = 1.0~\mathrm{GHz}$, $\epsilon_0=1$, $\epsilon_1=\epsilon_3=2.22$, $d_1=0.2~ \mu\mathrm{m}$, and $d_2=5~\mu\mathrm{m}$.}
\label{fig4}
\end{figure}
\begin{figure*}
\begin{center}
\includegraphics[width=5.5cm]{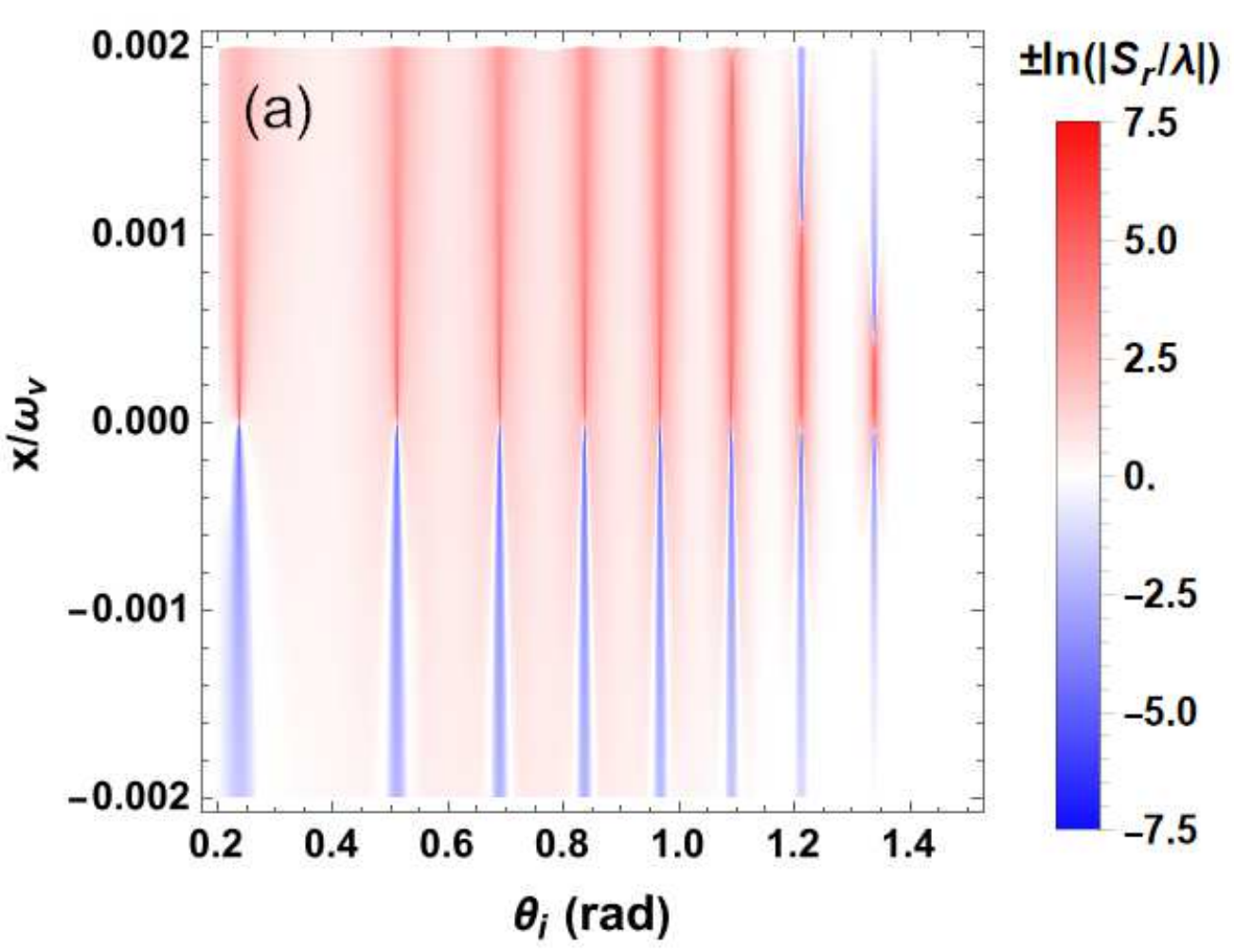}
\includegraphics[width=5.5cm]{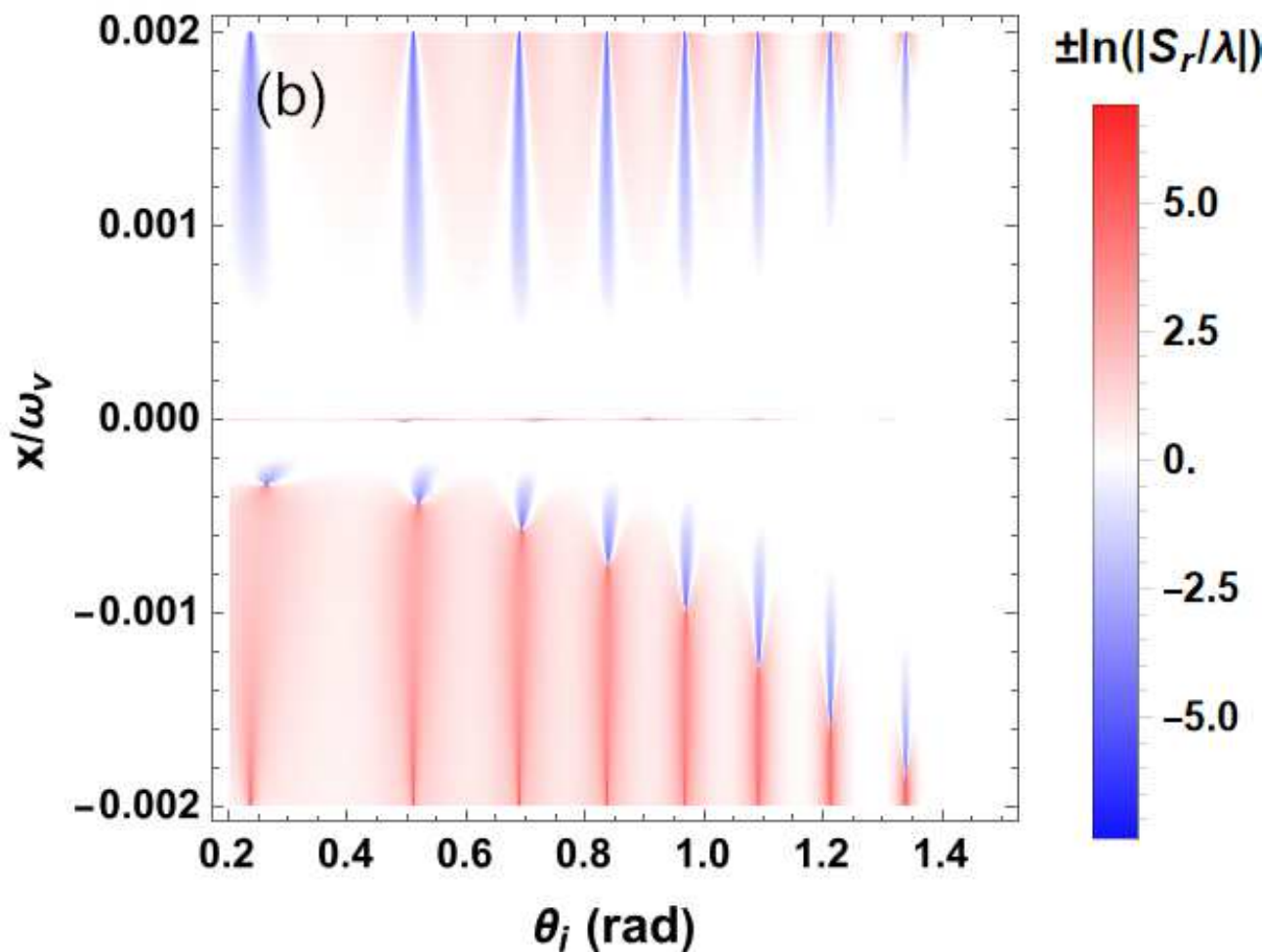}
\includegraphics[width=5.5cm]{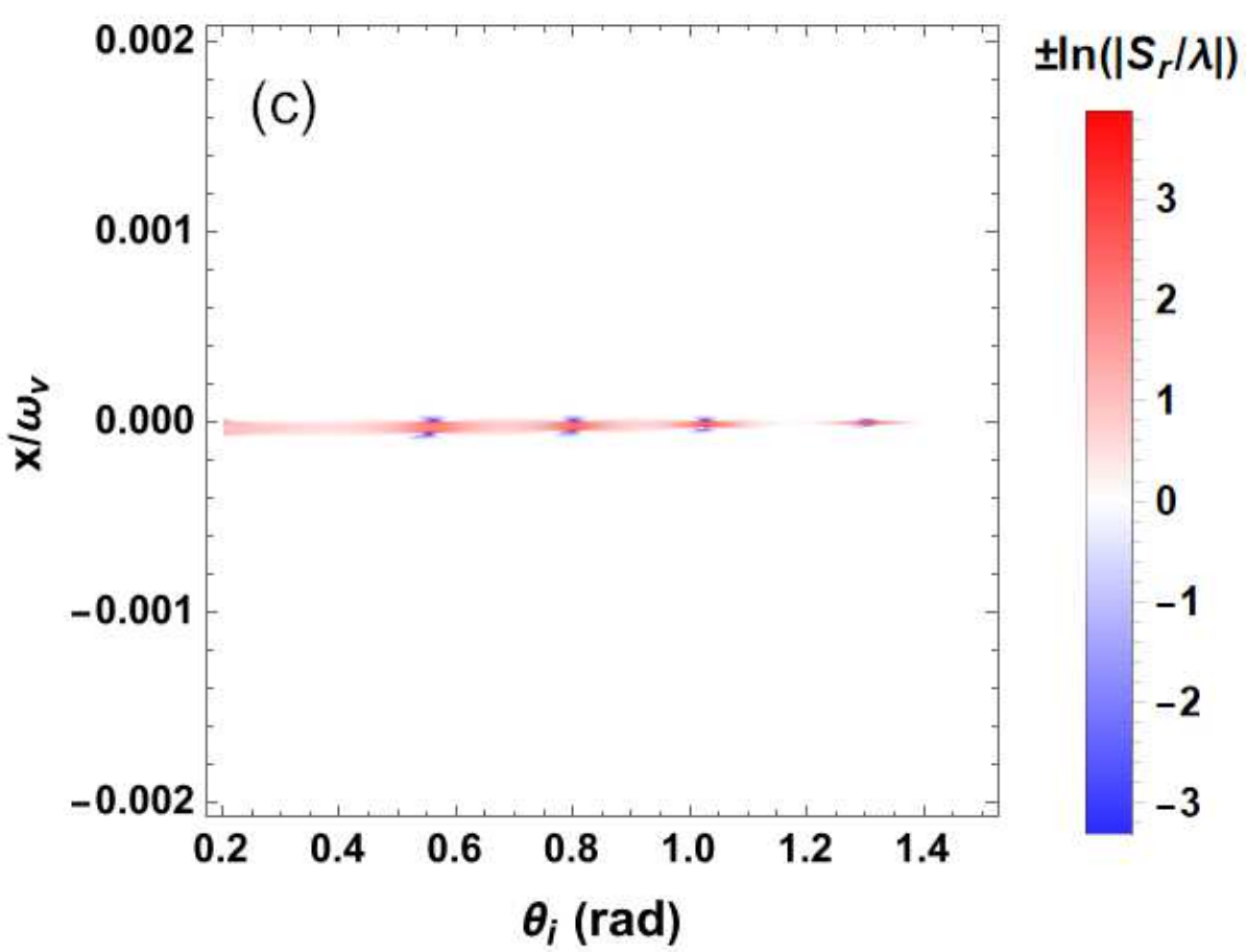}
\caption{Normalized GHS $S_r / \lambda$ as functions of effective detuning $x / \omega_v$ and probe field incident angle $\theta_i$ for effective exciton-vibration coupling (a) $G_v = 0$, (b) $G_v = 2\pi \times 0.06~\mathrm{THz}$, and (c) $G_v = 2\pi \times 0.3~\mathrm{THz}$. Fixed parameters: $\omega_a / 2\pi = 300~\mathrm{THz}$, $\omega_v / 2\pi = 30~\mathrm{THz}$, $G_a / 2\pi = 1.0~\mathrm{THz}$, $\kappa_a / 2\pi = \kappa_b / 2\pi = 0.2~\mathrm{THz}$, $\gamma / 2\pi = 1.0~\mathrm{GHz}$, $\epsilon_0 = 1$, $\epsilon_1 = \epsilon_3 = 2.22$, $d_1 = 0.2~\mu\mathrm{m}$, and $d_2 = 5~\mu\mathrm{m}$.}
\label{fig5}
\end{center}
\end{figure*}
In this section, we present numerical results using parameters from experimentally realizable systems~\cite{Shishkov-PRL, Shishkov-Room, Mart-PRL, Roelli2016, Schmidt2016, Carlon-PRL}: $\omega_a/2\pi = 300~\mathrm{THz}$, $\omega_v/2\pi = 30~\mathrm{THz}$, $\kappa_a/2\pi = \kappa_b/2\pi = 0.2~\mathrm{THz}$, $G_a/2\pi = 1.0~\mathrm{THz}$, $\gamma/2\pi = 1.0~\mathrm{GHz}$, and $N = 10000$. For the analysis of the GHS, we choose the dielectric parameters as $\epsilon_0=1$, $\epsilon_1=\epsilon_3=2.22$, $d_1=0.2~\mu\mathrm{m}$, and $d_2=5~\mu \mathrm{m}$~\cite{Zia_2010_Coherent, Wang_2008}. 

We begin our numerical analysis by examining the optical susceptibility (output probe field $E_T$) of the intermediate (intracavity) layer in the considered polariton optomechanical system, as given in Eq.~(\ref{optical sus}). Figure~\ref{fig2}(a) shows the output absorption spectrum $\mathrm{Re}[E_T]$ as a function of the normalized effective detuning $x/\omega_v$. In the absence of effective exciton–vibration coupling strength ($G_{v}=0$, solid blue), where the excitonic mode couples only to the optical mode, the output spectrum of the probe field shows two Lorentzian peaks with a single transparency dip at resonance $x=0$, a feature known as exciton-mediated OMIT. In the presence of the effective exciton–vibration interaction ($G_{v}=2\pi\times0.3~\mathrm{THz}$, dashed blue), the single exciton-mediated OMIT window splits into a double-transparency window known as exciton-vibration OMIT. This splitting arises from the nonzero molecular vibration interactions [Fig.~\ref{fig2}(a)]. Figure~\ref{fig2}(b) compares the dispersion spectrum  $\mathrm{Im}[E_T]$ of the probe field in the absence ($G_{v}=0$, solid blue) and presence ($G_{v}=2\pi\times0.3~\mathrm{THz}$, dashed blue) of effective exciton-vibration coupling. The width of the transparency window depends on the coupling strength between the collective excitonic and optical modes $G_{a}$.

In order to reveal the influence of effective exciton–vibration coupling $G_v$ on the absolute reflection coefficient $|r(k_{z}, \omega_{p})|$ and the normalized GHS $S_{r}/\lambda$, we present $|r(k_{z}, \omega_{p})|$ and $S_{r}/\lambda$ as a function of probe field incident angle $\theta_i$ for various values of $G_v$. The steady-state values [Eq.~\ref{steady states}] clearly indicate that all excitation modes are strongly dependent on the pump field strength through $b_s$. Furthermore, $G_{v}$ is directly proportional to $b_{s}$ for a fixed $g_{v}$. Consequently, the probe field spectrum can be effectively controlled by tuning the pump field strength via its effect on $G_v$. The reflection dips in Fig.~\ref{fig3}(a) signify resonance conditions, at which the GHS exhibits sharp positive peaks [Fig.~\ref{fig3}(b)]. For $G_v=0$ (green curves), these pronounced peaks result from steep phase dispersion at narrow exciton-photon polariton resonances. Introducing a finite effective exciton–vibration coupling strength through a pump field $G_v=2\pi\times0.06~\mathrm{THz}$ (blue curves) and  $G_v=2\pi\times0.3~\mathrm{THz}$ (red curves) broadens these resonances via vibrational dressing, which opens additional absorption/decay channels. This broadening reduces the slope of the reflection phase, thereby suppressing the amplitude of the positive GHS [Figs.~\ref{fig3}(a) and (b)]. 

To further elucidate the behavior associated with the effective exciton–vibration coupling strength $G_v$, we analyze the normalized GHS $S_{r}/\lambda$ as functions of $G_{v}$ and the incident angle of the probe field $\theta_{i}$. Figure~\ref{fig4} shows a contour plot of $S_{r}/\lambda$ versus $G_{v}$ and $\theta_{i}$ at the resonance condition ($x=0$). A pronounced GHS is observed at low $G_v$ values and larger $\theta_{i}$. However, as $G_v$ increases, the GHS is progressively suppressed. This suppression is attributed to enhanced absorption of the incident probe field at higher $G_v$ values, as inferred from the reflection coefficient [Eq.~\eqref{2x2}], which reduces the magnitude of the GHS. Overall, the trends displayed in this contour plot are consistent with the behaviors shown in Figs.~\ref{fig3}(a) and (b).
\begin{figure}
\begin{center}
\includegraphics[width=\linewidth]{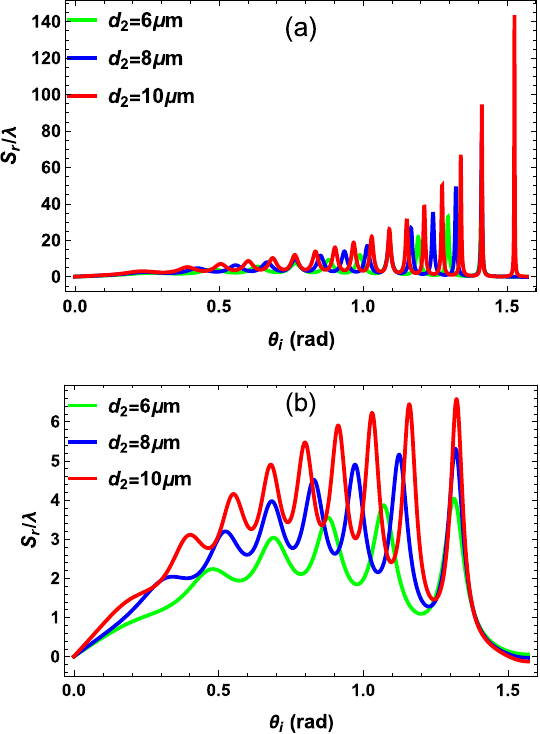}
\caption{Normalized GHS $S_r / \lambda$ as a function of probe field incident angle $\theta_i$ at resonance ($x=0$) for different intracavity lengths and effective exciton-vibration coupling (a) $G_v = 0$, and (b) $G_v = 2\pi \times 0.3~\mathrm{THz}$. Green, blue, and red curves correspond to $d_2 = 6~\mu\mathrm{m}$, $8~\mu\mathrm{m}$, and $10~\mu\mathrm{m}$, respectively. Fixed parameters: $\omega_a / 2\pi = 300~\mathrm{THz}$, $\omega_v / 2\pi = 30~\mathrm{THz}$, $G_a / 2\pi = 1.0~\mathrm{THz}$, $\kappa_a / 2\pi = \kappa_b / 2\pi = 0.2~\mathrm{THz}$, $\gamma / 2\pi = 1.0~\mathrm{GHz}$, $\epsilon_0 = 1$, $\epsilon_1 = \epsilon_3 = 2.22$, and $d_1 = 0.2~\mu\mathrm{m}$.}
\label{fig6}
\end{center}
\end{figure}
\begin{figure}
\begin{center}
\includegraphics[width=\linewidth]{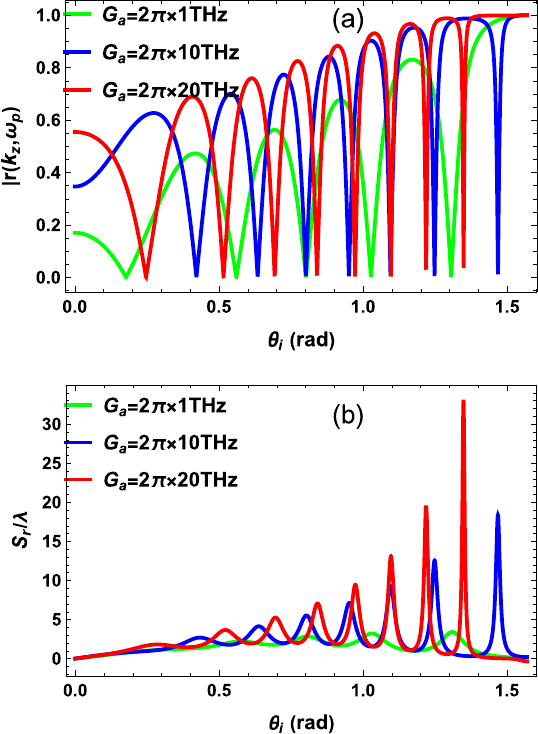}
\caption{(a) Absolute value of the reflection coefficient $|r(k_z, \omega_p)|$ and (b) normalized GHS $S_r / \lambda$ as a function of incident angle $\theta_i$ for various collective exciton-optical coupling strengths $G_a = 2\pi \times 1~\mathrm{THz}$ (green), $2\pi \times 10~\mathrm{THz}$ (blue), and $2\pi \times 20~\mathrm{THz}$ (red) at resonance ($x=0$). Fixed parameters: $\omega_a / 2\pi = 300~\mathrm{THz}$, $\omega_v / 2\pi = 30~\mathrm{THz}$, $G_v / 2\pi = 0.3~\mathrm{THz}$, $\kappa_a / 2\pi = \kappa_b / 2\pi = 0.2~\mathrm{THz}$, $\gamma / 2\pi = 1.0~\mathrm{GHz}$, $\epsilon_0 = 1$, $\epsilon_1 = \epsilon_3 = 2.22$, $d_1 = 0.2~\mu\mathrm{m}$, and $d_2 = 5~\mu\mathrm{m}$.}
\label{fig7}
\end{center}
\end{figure}

Next, we analyze the contour plots of the normalized GHS $S_{r}/\lambda$ as functions of the normalized effective detuning $x/\omega_{v}$ and probe field incident angle $\theta_{i}$, both in the absence and presence of the effective exciton–vibration coupling strength $G_{v}$. Distinct behaviors emerge in each case. At $G_{v} = 0$, the GHS exhibits a sharp transition from positive to negative peaks. Away from resonance, the magnitude of the GHS is enhanced at finite detuning and gradually diminishes at larger detuning values. This transition from positive to negative shifts is driven solely by the effective detuning. These features yield pronounced symmetric positive and negative GHS peaks at the resonance condition $x=0$ [Fig.~\ref{fig5}(a)]. In contrast, when the effective exciton–vibration coupling is introduced $G_{v}=2\pi\times0.06~\mathrm{THz}$, the GHS exhibits an asymmetric dispersion profile, with peaks shifted toward negative effective detuning [Fig.~\ref{fig5}(b)]. Consequently, the GHS remains in the negative regime over a wider detuning range. At $G_{v}= 2\pi\times0.3~\mathrm{THz}$ the GHS is suppressed [Fig.~\ref{fig5}(c)].

The observed sign reversal of the GHS can be understood in terms of the group index $N_g$ of the cavity system~\cite{Zia_2010_Coherent, Wild-1982}. It can be approximated by the phase derivative with respect to angular frequency $N_g \approx {1}/{L}[{d\phi_r}/{d\omega_p}]$, where $L = 2d_1 + d_2$ is the total thickness of the cavity system. In this framework, the sign of the GHS is correlated with the sign of the effective group index~\cite{Zia_2010_Coherent}. These findings directly explain the trend in Figs.~\ref{fig5}(a-c), linking the detuning- and effective exciton-vibration coupling-dependent behavior of the GHS to the underlying dispersion properties of the cavity.

Cavity dimensions serve as another crucial control parameter for the GHS magnitude. We therefore examine the influence of cavity geometry on the GHS. Our results reveal that the GHS is highly sensitive to the cavity structure, particularly to the length of the intracavity medium $d_{2}$. To quantify this effect, we show the GHS as a function of the probe field incident angle $\theta_{i}$ for three representative values of $d_{2}$ (i.e., $6~\mu\mathrm{m}$, $8~\mu\mathrm{m}$, and $10~\mu\mathrm{m}$). Figure~\ref{fig6}(a) illustrates the case with $G_{v}=0$, whereas Fig.~\ref{fig6}(b) corresponds to $G_{v}=2\pi \times 0.3~\mathrm{THz}$. A clear trend emerges whereby increasing the intracavity length enhances both the number of resonance peaks and the overall magnitude of the GHS in both cases. This thickness-dependent modulation of the GHS in polariton optomechanics is consistent with earlier findings in atomic media in cavity~\cite{Zia_2010_Coherent} and cavity magnomechanical systems~\cite{waseem_Goos_2024}.

We further investigate the influence of the collective exciton–optical coupling strength $G_{a}$ on the reflection coefficient $|r(k_{z}, \omega_{p})|$ and the associated GHS as a function of the probe field incident angle $\theta_i$, while keeping the effective exciton–vibration coupling fixed at $G_{v}=2\pi\times0.3~\mathrm{THz}$. Figures~\ref{fig7}(a) and (b) show $|r(k_{z}, \omega_{p})|$ and the normalized GHS $S_{r}/\lambda$, respectively, versus $\theta_i$. 
Increasing $G_{a}$ sharpens the resonance dips in the reflection spectrum [Fig.~\ref{fig7}(a)], which results in a corresponding increase in the magnitude of the GHS [Fig.~\ref{fig7}(b)]. This enhancement reflects a deeper penetration of the probe field into the cavity, driven by the intensified light-matter interaction.

\section{Conclusion}
In summary, we have demonstrated a theoretical framework for the coherent control of the GHS in a polariton optomechanical system. Our key finding is that the effective exciton–vibration coupling acts as a dynamic switch: in its absence, the system exhibits exciton-mediated OMIT, whereas its activation converts the response into exciton–vibration OMIT. A pronounced GHS at the resonance condition is strongly suppressed once this effective coupling is turned on. Furthermore, the effective exciton–vibration coupling fundamentally alters the system’s symmetry: without it, the GHS varies symmetrically with respect to the effective cavity detuning, while its presence breaks this symmetry and produces substantial shifts over a broad range of negative detunings. The GHS can also be tuned by adjusting the cavity length, and increasing the collective exciton–optical coupling strength enhances the GHS magnitude for a fixed effective exciton–vibration coupling.

These results establish a versatile platform for exploring and manipulating light-matter interactions via the GHS in hybrid quantum systems, paving the way for innovative optical devices that leverage beam-displacement phenomena for advanced applications in sensing and quantum information processing.

\section*{Acknowledgement}
We acknowledge the financial support from the NSFC under grant 
No. 12174346.\\
\\

\section*{references}

\bibliography{ref} 

\begin{thebibliography}{72}%
\makeatletter
\providecommand \@ifxundefined [1]{%
 \@ifx{#1\undefined}
}%
\providecommand \@ifnum [1]{%
 \ifnum #1\expandafter \@firstoftwo
 \else \expandafter \@secondoftwo
 \fi
}%
\providecommand \@ifx [1]{%
 \ifx #1\expandafter \@firstoftwo
 \else \expandafter \@secondoftwo
 \fi
}%
\providecommand \natexlab [1]{#1}%
\providecommand \enquote  [1]{``#1''}%
\providecommand \bibnamefont  [1]{#1}%
\providecommand \bibfnamefont [1]{#1}%
\providecommand \citenamefont [1]{#1}%
\providecommand \href@noop [0]{\@secondoftwo}%
\providecommand \href [0]{\begingroup \@sanitize@url \@href}%
\providecommand \@href[1]{\@@startlink{#1}\@@href}%
\providecommand \@@href[1]{\endgroup#1\@@endlink}%
\providecommand \@sanitize@url [0]{\catcode `\\12\catcode `\$12\catcode `\&12\catcode `\#12\catcode `\^12\catcode `\_12\catcode `\%12\relax}%
\providecommand \@@startlink[1]{}%
\providecommand \@@endlink[0]{}%
\providecommand \url  [0]{\begingroup\@sanitize@url \@url }%
\providecommand \@url [1]{\endgroup\@href {#1}{\urlprefix }}%
\providecommand \urlprefix  [0]{URL }%
\providecommand \Eprint [0]{\href }%
\providecommand \doibase [0]{http://dx.doi.org/}%
\providecommand \selectlanguage [0]{\@gobble}%
\providecommand \bibinfo  [0]{\@secondoftwo}%
\providecommand \bibfield  [0]{\@secondoftwo}%
\providecommand \translation [1]{[#1]}%
\providecommand \BibitemOpen [0]{}%
\providecommand \bibitemStop [0]{}%
\providecommand \bibitemNoStop [0]{.\EOS\space}%
\providecommand \EOS [0]{\spacefactor3000\relax}%
\providecommand \BibitemShut  [1]{\csname bibitem#1\endcsname}%
\let\auto@bib@innerbib\@empty
\bibitem [{\citenamefont {Aspelmeyer}\ \emph {et~al.}(2014{\natexlab{a}})\citenamefont {Aspelmeyer}, \citenamefont {Kippenberg},\ and\ \citenamefont {Marquardt}}]{Aspelmeyer-book-2014}%
  \BibitemOpen
  \bibfield  {author} {\bibinfo {author} {\bibfnamefont {Markus}\ \bibnamefont {Aspelmeyer}}, \bibinfo {author} {\bibfnamefont {Tobias~J.}\ \bibnamefont {Kippenberg}}, \ and\ \bibinfo {author} {\bibfnamefont {Florian}\ \bibnamefont {Marquardt}},\ }\href@noop {} {\emph {\bibinfo {title} {Cavity Optomechanics: Nano and Micromechanical Resonators Interacting with Light}}}\ (\bibinfo  {publisher} {Springer},\ \bibinfo {address} {Berlin, Heidelberg},\ \bibinfo {year} {2014})\BibitemShut {NoStop}%
\bibitem [{\citenamefont {Aspelmeyer}\ \emph {et~al.}(2014{\natexlab{b}})\citenamefont {Aspelmeyer}, \citenamefont {Kippenberg},\ and\ \citenamefont {Marquardt}}]{Aspelmeyer2014}%
  \BibitemOpen
  \bibfield  {author} {\bibinfo {author} {\bibfnamefont {Markus}\ \bibnamefont {Aspelmeyer}}, \bibinfo {author} {\bibfnamefont {Tobias~J.}\ \bibnamefont {Kippenberg}}, \ and\ \bibinfo {author} {\bibfnamefont {Florian}\ \bibnamefont {Marquardt}},\ }\bibfield  {title} {\enquote {\bibinfo {title} {Cavity optomechanics},}\ }\href {\doibase 10.1103/RevModPhys.86.1391} {\bibfield  {journal} {\bibinfo  {journal} {Rev. Mod. Phys.}\ }\textbf {\bibinfo {volume} {86}},\ \bibinfo {pages} {1391} (\bibinfo {year} {2014}{\natexlab{b}})}\BibitemShut {NoStop}%
\bibitem [{\citenamefont {Kippenberg}\ and\ \citenamefont {Vahala}(2008)}]{Kippenberg}%
  \BibitemOpen
  \bibfield  {author} {\bibinfo {author} {\bibfnamefont {T.~J.}\ \bibnamefont {Kippenberg}}\ and\ \bibinfo {author} {\bibfnamefont {K.~J.}\ \bibnamefont {Vahala}},\ }\bibfield  {title} {\enquote {\bibinfo {title} {Cavity optomechanics: Back-action at the mesoscale},}\ }\href {\doibase 10.1126/science.1156032} {\bibfield  {journal} {\bibinfo  {journal} {Science}\ }\textbf {\bibinfo {volume} {321}},\ \bibinfo {pages} {1172} (\bibinfo {year} {2008})}\BibitemShut {NoStop}%
\bibitem [{\citenamefont {Metcalfe}(2014)}]{Metcalfe-2014}%
  \BibitemOpen
  \bibfield  {author} {\bibinfo {author} {\bibfnamefont {Michael}\ \bibnamefont {Metcalfe}},\ }\bibfield  {title} {\enquote {\bibinfo {title} {Applications of cavity optomechanics},}\ }\href {\doibase 10.1063/1.4896029} {\bibfield  {journal} {\bibinfo  {journal} {Appl. Phys. Rev.}\ }\textbf {\bibinfo {volume} {1}},\ \bibinfo {pages} {031105} (\bibinfo {year} {2014})}\BibitemShut {NoStop}%
\bibitem [{\citenamefont {Barzanjeh}\ \emph {et~al.}(2022)\citenamefont {Barzanjeh}, \citenamefont {Xuereb}, \citenamefont {Gr{\"o}blacher}, \citenamefont {Paternostro}, \citenamefont {Regal},\ and\ \citenamefont {Weig}}]{Barzanjeh2022}%
  \BibitemOpen
  \bibfield  {author} {\bibinfo {author} {\bibfnamefont {Shabir}\ \bibnamefont {Barzanjeh}}, \bibinfo {author} {\bibfnamefont {Andr{\'e}}\ \bibnamefont {Xuereb}}, \bibinfo {author} {\bibfnamefont {Simon}\ \bibnamefont {Gr{\"o}blacher}}, \bibinfo {author} {\bibfnamefont {Mauro}\ \bibnamefont {Paternostro}}, \bibinfo {author} {\bibfnamefont {Cindy~A.}\ \bibnamefont {Regal}}, \ and\ \bibinfo {author} {\bibfnamefont {Eva~M.}\ \bibnamefont {Weig}},\ }\bibfield  {title} {{\selectlanguage {English}\enquote {\bibinfo {title} {Optomechanics for quantum technologies},}\ }}\href {\doibase 10.1038/s41567-021-01402-0} {\bibfield  {journal} {\bibinfo  {journal} {Nat. Phys.}\ }\textbf {\bibinfo {volume} {18}},\ \bibinfo {pages} {15} (\bibinfo {year} {2022})}\BibitemShut {NoStop}%
\bibitem [{\citenamefont {O'Connell}\ \emph {et~al.}(2010)\citenamefont {O'Connell}, \citenamefont {Hofheinz}, \citenamefont {Ansmann}, \citenamefont {Bialczak}, \citenamefont {Lenander}, \citenamefont {Lucero}, \citenamefont {Neeley}, \citenamefont {Sank}, \citenamefont {Wang}, \citenamefont {Weides}, \citenamefont {Wenner}, \citenamefont {Martinis},\ and\ \citenamefont {Cleland}}]{Connell2010}%
  \BibitemOpen
  \bibfield  {author} {\bibinfo {author} {\bibfnamefont {A.~D.}\ \bibnamefont {O'Connell}}, \bibinfo {author} {\bibfnamefont {M.}~\bibnamefont {Hofheinz}}, \bibinfo {author} {\bibfnamefont {M.}~\bibnamefont {Ansmann}}, \bibinfo {author} {\bibfnamefont {Radoslaw~C.}\ \bibnamefont {Bialczak}}, \bibinfo {author} {\bibfnamefont {M.}~\bibnamefont {Lenander}}, \bibinfo {author} {\bibfnamefont {Erik}\ \bibnamefont {Lucero}}, \bibinfo {author} {\bibfnamefont {M.}~\bibnamefont {Neeley}}, \bibinfo {author} {\bibfnamefont {D.}~\bibnamefont {Sank}}, \bibinfo {author} {\bibfnamefont {H.}~\bibnamefont {Wang}}, \bibinfo {author} {\bibfnamefont {M.}~\bibnamefont {Weides}}, \bibinfo {author} {\bibfnamefont {J.}~\bibnamefont {Wenner}}, \bibinfo {author} {\bibfnamefont {John~M.}\ \bibnamefont {Martinis}}, \ and\ \bibinfo {author} {\bibfnamefont {A.~N.}\ \bibnamefont {Cleland}},\ }\bibfield  {title} {\enquote {\bibinfo {title} {Quantum ground state and single-phonon control of a mechanical resonator},}\ }\href {\doibase 10.1038/nature08967} {\bibfield  {journal} {\bibinfo  {journal} {Nature}\ }\textbf {\bibinfo {volume} {464}},\ \bibinfo {pages} {697} (\bibinfo {year} {2010})}\BibitemShut {NoStop}%
\bibitem [{\citenamefont {Gustafsson}\ \emph {et~al.}(2014)\citenamefont {Gustafsson}, \citenamefont {Aref}, \citenamefont {Kockum}, \citenamefont {Ekström}, \citenamefont {Johansson},\ and\ \citenamefont {Delsing}}]{Martin-2017}%
  \BibitemOpen
  \bibfield  {author} {\bibinfo {author} {\bibfnamefont {Martin~V.}\ \bibnamefont {Gustafsson}}, \bibinfo {author} {\bibfnamefont {Thomas}\ \bibnamefont {Aref}}, \bibinfo {author} {\bibfnamefont {Anton~Frisk}\ \bibnamefont {Kockum}}, \bibinfo {author} {\bibfnamefont {Maria~K.}\ \bibnamefont {Ekström}}, \bibinfo {author} {\bibfnamefont {Göran}\ \bibnamefont {Johansson}}, \ and\ \bibinfo {author} {\bibfnamefont {Per}\ \bibnamefont {Delsing}},\ }\bibfield  {title} {\enquote {\bibinfo {title} {Propagating phonons coupled to an artificial atom},}\ }\href {\doibase 10.1126/science.1257219} {\bibfield  {journal} {\bibinfo  {journal} {Science}\ }\textbf {\bibinfo {volume} {346}},\ \bibinfo {pages} {207} (\bibinfo {year} {2014})}\BibitemShut {NoStop}%
\bibitem [{\citenamefont {Chu}\ \emph {et~al.}(2017)\citenamefont {Chu}, \citenamefont {Kharel}, \citenamefont {Renninger}, \citenamefont {Burkhart}, \citenamefont {Frunzio}, \citenamefont {Rakich},\ and\ \citenamefont {Schoelkopf}}]{Yiwen-2017}%
  \BibitemOpen
  \bibfield  {author} {\bibinfo {author} {\bibfnamefont {Yiwen}\ \bibnamefont {Chu}}, \bibinfo {author} {\bibfnamefont {Prashanta}\ \bibnamefont {Kharel}}, \bibinfo {author} {\bibfnamefont {William~H.}\ \bibnamefont {Renninger}}, \bibinfo {author} {\bibfnamefont {Luke~D.}\ \bibnamefont {Burkhart}}, \bibinfo {author} {\bibfnamefont {Luigi}\ \bibnamefont {Frunzio}}, \bibinfo {author} {\bibfnamefont {Peter~T.}\ \bibnamefont {Rakich}}, \ and\ \bibinfo {author} {\bibfnamefont {Robert~J.}\ \bibnamefont {Schoelkopf}},\ }\bibfield  {title} {\enquote {\bibinfo {title} {Quantum acoustics with superconducting qubits},}\ }\href {\doibase 10.1126/science.aao1511} {\bibfield  {journal} {\bibinfo  {journal} {Science}\ }\textbf {\bibinfo {volume} {358}},\ \bibinfo {pages} {199} (\bibinfo {year} {2017})}\BibitemShut {NoStop}%
\bibitem [{\citenamefont {Mancini}\ \emph {et~al.}(2002)\citenamefont {Mancini}, \citenamefont {Giovannetti}, \citenamefont {Vitali},\ and\ \citenamefont {Tombesi}}]{Mancini-2002}%
  \BibitemOpen
  \bibfield  {author} {\bibinfo {author} {\bibfnamefont {Stefano}\ \bibnamefont {Mancini}}, \bibinfo {author} {\bibfnamefont {Vittorio}\ \bibnamefont {Giovannetti}}, \bibinfo {author} {\bibfnamefont {David}\ \bibnamefont {Vitali}}, \ and\ \bibinfo {author} {\bibfnamefont {Paolo}\ \bibnamefont {Tombesi}},\ }\bibfield  {title} {\enquote {\bibinfo {title} {Entangling macroscopic oscillators exploiting radiation pressure},}\ }\href {\doibase 10.1103/PhysRevLett.88.120401} {\bibfield  {journal} {\bibinfo  {journal} {Phys. Rev. Lett.}\ }\textbf {\bibinfo {volume} {88}},\ \bibinfo {pages} {120401} (\bibinfo {year} {2002})}\BibitemShut {NoStop}%
\bibitem [{\citenamefont {Vitali}\ \emph {et~al.}(2007)\citenamefont {Vitali}, \citenamefont {Gigan}, \citenamefont {Ferreira}, \citenamefont {B\"ohm}, \citenamefont {Tombesi}, \citenamefont {Guerreiro}, \citenamefont {Vedral}, \citenamefont {Zeilinger},\ and\ \citenamefont {Aspelmeyer}}]{Vitali-2007}%
  \BibitemOpen
  \bibfield  {author} {\bibinfo {author} {\bibfnamefont {D.}~\bibnamefont {Vitali}}, \bibinfo {author} {\bibfnamefont {S.}~\bibnamefont {Gigan}}, \bibinfo {author} {\bibfnamefont {A.}~\bibnamefont {Ferreira}}, \bibinfo {author} {\bibfnamefont {H.~R.}\ \bibnamefont {B\"ohm}}, \bibinfo {author} {\bibfnamefont {P.}~\bibnamefont {Tombesi}}, \bibinfo {author} {\bibfnamefont {A.}~\bibnamefont {Guerreiro}}, \bibinfo {author} {\bibfnamefont {V.}~\bibnamefont {Vedral}}, \bibinfo {author} {\bibfnamefont {A.}~\bibnamefont {Zeilinger}}, \ and\ \bibinfo {author} {\bibfnamefont {M.}~\bibnamefont {Aspelmeyer}},\ }\bibfield  {title} {\enquote {\bibinfo {title} {Optomechanical entanglement between a movable mirror and a cavity field},}\ }\href {\doibase 10.1103/PhysRevLett.98.030405} {\bibfield  {journal} {\bibinfo  {journal} {Phys. Rev. Lett.}\ }\textbf {\bibinfo {volume} {98}},\ \bibinfo {pages} {030405} (\bibinfo {year} {2007})}\BibitemShut {NoStop}%
\bibitem [{\citenamefont {Paternostro}\ \emph {et~al.}(2007)\citenamefont {Paternostro}, \citenamefont {Vitali}, \citenamefont {Gigan}, \citenamefont {Kim}, \citenamefont {Brukner}, \citenamefont {Eisert},\ and\ \citenamefont {Aspelmeyer}}]{Paternostro-2007}%
  \BibitemOpen
  \bibfield  {author} {\bibinfo {author} {\bibfnamefont {M.}~\bibnamefont {Paternostro}}, \bibinfo {author} {\bibfnamefont {D.}~\bibnamefont {Vitali}}, \bibinfo {author} {\bibfnamefont {S.}~\bibnamefont {Gigan}}, \bibinfo {author} {\bibfnamefont {M.~S.}\ \bibnamefont {Kim}}, \bibinfo {author} {\bibfnamefont {C.}~\bibnamefont {Brukner}}, \bibinfo {author} {\bibfnamefont {J.}~\bibnamefont {Eisert}}, \ and\ \bibinfo {author} {\bibfnamefont {M.}~\bibnamefont {Aspelmeyer}},\ }\bibfield  {title} {\enquote {\bibinfo {title} {Creating and probing multipartite macroscopic entanglement with light},}\ }\href {\doibase 10.1103/PhysRevLett.99.250401} {\bibfield  {journal} {\bibinfo  {journal} {Phys. Rev. Lett.}\ }\textbf {\bibinfo {volume} {99}},\ \bibinfo {pages} {250401} (\bibinfo {year} {2007})}\BibitemShut {NoStop}%
\bibitem [{\citenamefont {Wilson-Rae}\ \emph {et~al.}(2007)\citenamefont {Wilson-Rae}, \citenamefont {Nooshi}, \citenamefont {Zwerger},\ and\ \citenamefont {Kippenberg}}]{Wilson-PRL}%
  \BibitemOpen
  \bibfield  {author} {\bibinfo {author} {\bibfnamefont {I.}~\bibnamefont {Wilson-Rae}}, \bibinfo {author} {\bibfnamefont {N.}~\bibnamefont {Nooshi}}, \bibinfo {author} {\bibfnamefont {W.}~\bibnamefont {Zwerger}}, \ and\ \bibinfo {author} {\bibfnamefont {T.~J.}\ \bibnamefont {Kippenberg}},\ }\bibfield  {title} {\enquote {\bibinfo {title} {Theory of ground state cooling of a mechanical oscillator using dynamical backaction},}\ }\href {\doibase 10.1103/PhysRevLett.99.093901} {\bibfield  {journal} {\bibinfo  {journal} {Phys. Rev. Lett.}\ }\textbf {\bibinfo {volume} {99}},\ \bibinfo {pages} {093901} (\bibinfo {year} {2007})}\BibitemShut {NoStop}%
\bibitem [{\citenamefont {Marquardt}\ \emph {et~al.}(2007)\citenamefont {Marquardt}, \citenamefont {Chen}, \citenamefont {Clerk},\ and\ \citenamefont {Girvin}}]{Marquardt-PRL}%
  \BibitemOpen
  \bibfield  {author} {\bibinfo {author} {\bibfnamefont {Florian}\ \bibnamefont {Marquardt}}, \bibinfo {author} {\bibfnamefont {Joe~P.}\ \bibnamefont {Chen}}, \bibinfo {author} {\bibfnamefont {A.~A.}\ \bibnamefont {Clerk}}, \ and\ \bibinfo {author} {\bibfnamefont {S.~M.}\ \bibnamefont {Girvin}},\ }\bibfield  {title} {\enquote {\bibinfo {title} {{Quantum Theory of Cavity-Assisted Sideband Cooling of Mechanical Motion}},}\ }\href {\doibase 10.1103/PhysRevLett.99.093902} {\bibfield  {journal} {\bibinfo  {journal} {Phys. Rev. Lett.}\ }\textbf {\bibinfo {volume} {99}},\ \bibinfo {pages} {093902} (\bibinfo {year} {2007})}\BibitemShut {NoStop}%
\bibitem [{\citenamefont {Chan}\ \emph {et~al.}(2011)\citenamefont {Chan}, \citenamefont {Alegre}, \citenamefont {Safavi-Naeini}, \citenamefont {Hill}, \citenamefont {Krause}, \citenamefont {Gröblacher}, \citenamefont {Aspelmeyer},\ and\ \citenamefont {Painter}}]{Jasper-2011}%
  \BibitemOpen
  \bibfield  {author} {\bibinfo {author} {\bibfnamefont {Jasper}\ \bibnamefont {Chan}}, \bibinfo {author} {\bibfnamefont {T.~P.~Mayer}\ \bibnamefont {Alegre}}, \bibinfo {author} {\bibfnamefont {Amir~H.}\ \bibnamefont {Safavi-Naeini}}, \bibinfo {author} {\bibfnamefont {Jeff~T.}\ \bibnamefont {Hill}}, \bibinfo {author} {\bibfnamefont {Alex}\ \bibnamefont {Krause}}, \bibinfo {author} {\bibfnamefont {Simon}\ \bibnamefont {Gröblacher}}, \bibinfo {author} {\bibfnamefont {Markus}\ \bibnamefont {Aspelmeyer}}, \ and\ \bibinfo {author} {\bibfnamefont {Oskar}\ \bibnamefont {Painter}},\ }\bibfield  {title} {\enquote {\bibinfo {title} {Laser cooling of a nanomechanical oscillator into its quantum ground state},}\ }\href {\doibase 10.1038/nature10461} {\bibfield  {journal} {\bibinfo  {journal} {Nature}\ }\textbf {\bibinfo {volume} {478}},\ \bibinfo {pages} {89} (\bibinfo {year} {2011})}\BibitemShut {NoStop}%
\bibitem [{\citenamefont {Verhagen}\ \emph {et~al.}(2012)\citenamefont {Verhagen}, \citenamefont {Del{\'e}glise}, \citenamefont {Weis}, \citenamefont {Schliesser},\ and\ \citenamefont {Kippenberg}}]{Verhagen2012}%
  \BibitemOpen
  \bibfield  {author} {\bibinfo {author} {\bibfnamefont {E.}~\bibnamefont {Verhagen}}, \bibinfo {author} {\bibfnamefont {S.}~\bibnamefont {Del{\'e}glise}}, \bibinfo {author} {\bibfnamefont {S.}~\bibnamefont {Weis}}, \bibinfo {author} {\bibfnamefont {A.}~\bibnamefont {Schliesser}}, \ and\ \bibinfo {author} {\bibfnamefont {T.~J.}\ \bibnamefont {Kippenberg}},\ }\bibfield  {title} {\enquote {\bibinfo {title} {Quantum-coherent coupling of a mechanical oscillator to an optical cavity mode},}\ }\href {\doibase 10.1038/nature10787} {\bibfield  {journal} {\bibinfo  {journal} {Nature}\ }\textbf {\bibinfo {volume} {482}},\ \bibinfo {pages} {63} (\bibinfo {year} {2012})}\BibitemShut {NoStop}%
\bibitem [{\citenamefont {Weis}\ \emph {et~al.}(2010)\citenamefont {Weis}, \citenamefont {Rivière}, \citenamefont {Deléglise}, \citenamefont {Gavartin}, \citenamefont {Arcizet}, \citenamefont {Schliesser},\ and\ \citenamefont {Kippenberg}}]{Stefan-2010}%
  \BibitemOpen
  \bibfield  {author} {\bibinfo {author} {\bibfnamefont {Stefan}\ \bibnamefont {Weis}}, \bibinfo {author} {\bibfnamefont {Rémi}\ \bibnamefont {Rivière}}, \bibinfo {author} {\bibfnamefont {Samuel}\ \bibnamefont {Deléglise}}, \bibinfo {author} {\bibfnamefont {Emanuel}\ \bibnamefont {Gavartin}}, \bibinfo {author} {\bibfnamefont {Olivier}\ \bibnamefont {Arcizet}}, \bibinfo {author} {\bibfnamefont {Albert}\ \bibnamefont {Schliesser}}, \ and\ \bibinfo {author} {\bibfnamefont {Tobias~J.}\ \bibnamefont {Kippenberg}},\ }\bibfield  {title} {\enquote {\bibinfo {title} {Optomechanically induced transparency},}\ }\href {\doibase 10.1126/science.1195596} {\bibfield  {journal} {\bibinfo  {journal} {Science}\ }\textbf {\bibinfo {volume} {330}},\ \bibinfo {pages} {1520} (\bibinfo {year} {2010})}\BibitemShut {NoStop}%
\bibitem [{\citenamefont {Schliesser}(2009)}]{Schliesser-2009}%
  \BibitemOpen
  \bibfield  {author} {\bibinfo {author} {\bibfnamefont {Albert}\ \bibnamefont {Schliesser}},\ }\emph {\bibinfo {title} {Cavity Optomechanics and Optical Frequency Comb Generation with Silica Whispering-Gallery-Mode Microresonators}},\ \href {http://nbn-resolving.de/urn:nbn:de:bvb:19-109408} {Ph.D. thesis},\ \bibinfo  {school} {Ludwig-Maximilians-Universit{\"a}t M{\"u}nchen} (\bibinfo {year} {2009})\BibitemShut {NoStop}%
\bibitem [{\citenamefont {Schliesser}\ \emph {et~al.}(2010)\citenamefont {Schliesser}, \citenamefont {Kippenberg}, \citenamefont {Arimondo}, \citenamefont {Berman},\ and\ \citenamefont {Lin}}]{Schliesser-2010}%
  \BibitemOpen
  \bibfield  {author} {\bibinfo {author} {\bibfnamefont {Albert}\ \bibnamefont {Schliesser}}, \bibinfo {author} {\bibfnamefont {Tobias~J.}\ \bibnamefont {Kippenberg}}, \bibinfo {author} {\bibfnamefont {E.}~\bibnamefont {Arimondo}}, \bibinfo {author} {\bibfnamefont {P.~R.}\ \bibnamefont {Berman}}, \ and\ \bibinfo {author} {\bibfnamefont {C.~C.}\ \bibnamefont {Lin}},\ }\bibfield  {title} {{\selectlanguage {English}\enquote {\bibinfo {title} {Cavity optomechanics with whispering-gallery mode optical micro-resonators},}\ }}\href {\doibase 10.1016/S1049-250X(10)05810-6} {\bibfield  {journal} {\bibinfo  {journal} {Adv. At. Mol. Opt. Phys.}\ }\textbf {\bibinfo {volume} {58}},\ \bibinfo {pages} {207} (\bibinfo {year} {2010})}\BibitemShut {NoStop}%
\bibitem [{\citenamefont {Agarwal}\ and\ \citenamefont {Huang}(2010)}]{Agarwal-2010}%
  \BibitemOpen
  \bibfield  {author} {\bibinfo {author} {\bibfnamefont {G.~S.}\ \bibnamefont {Agarwal}}\ and\ \bibinfo {author} {\bibfnamefont {Sumei}\ \bibnamefont {Huang}},\ }\bibfield  {title} {\enquote {\bibinfo {title} {Electromagnetically induced transparency in mechanical effects of light},}\ }\href {\doibase 10.1103/PhysRevA.81.041803} {\bibfield  {journal} {\bibinfo  {journal} {Phys. Rev. A}\ }\textbf {\bibinfo {volume} {81}},\ \bibinfo {pages} {041803} (\bibinfo {year} {2010})}\BibitemShut {NoStop}%
\bibitem [{\citenamefont {Roelli}\ \emph {et~al.}(2016)\citenamefont {Roelli}, \citenamefont {Galland}, \citenamefont {Piro},\ and\ \citenamefont {Kippenberg}}]{Roelli2016}%
  \BibitemOpen
  \bibfield  {author} {\bibinfo {author} {\bibfnamefont {Philippe}\ \bibnamefont {Roelli}}, \bibinfo {author} {\bibfnamefont {Christophe}\ \bibnamefont {Galland}}, \bibinfo {author} {\bibfnamefont {Nicolas}\ \bibnamefont {Piro}}, \ and\ \bibinfo {author} {\bibfnamefont {Tobias~J.}\ \bibnamefont {Kippenberg}},\ }\bibfield  {title} {\enquote {\bibinfo {title} {Molecular cavity optomechanics as a theory of plasmon-enhanced raman scattering},}\ }\href {\doibase 10.1038/nnano.2015.264} {\bibfield  {journal} {\bibinfo  {journal} {Nat. Nanotechnol.}\ }\textbf {\bibinfo {volume} {11}},\ \bibinfo {pages} {164} (\bibinfo {year} {2016})}\BibitemShut {NoStop}%
\bibitem [{\citenamefont {Schmidt}\ \emph {et~al.}(2016)\citenamefont {Schmidt}, \citenamefont {Esteban}, \citenamefont {González-Tudela}, \citenamefont {Giedke},\ and\ \citenamefont {Aizpurua}}]{Schmidt2016}%
  \BibitemOpen
  \bibfield  {author} {\bibinfo {author} {\bibfnamefont {Mikolaj~K.}\ \bibnamefont {Schmidt}}, \bibinfo {author} {\bibfnamefont {Ruben}\ \bibnamefont {Esteban}}, \bibinfo {author} {\bibfnamefont {Alejandro}\ \bibnamefont {González-Tudela}}, \bibinfo {author} {\bibfnamefont {Geza}\ \bibnamefont {Giedke}}, \ and\ \bibinfo {author} {\bibfnamefont {Javier}\ \bibnamefont {Aizpurua}},\ }\bibfield  {title} {\enquote {\bibinfo {title} {{Quantum Mechanical Description of Raman Scattering from Molecules in Plasmonic Cavities}},}\ }\href {\doibase 10.1021/acsnano.6b02484} {\bibfield  {journal} {\bibinfo  {journal} {ACS Nano}\ }\textbf {\bibinfo {volume} {10}},\ \bibinfo {pages} {6291} (\bibinfo {year} {2016})}\BibitemShut {NoStop}%
\bibitem [{\citenamefont {Hughes}\ \emph {et~al.}(2021)\citenamefont {Hughes}, \citenamefont {Settineri}, \citenamefont {Savasta},\ and\ \citenamefont {Nori}}]{Hughes2021}%
  \BibitemOpen
  \bibfield  {author} {\bibinfo {author} {\bibfnamefont {Stephen}\ \bibnamefont {Hughes}}, \bibinfo {author} {\bibfnamefont {Alessio}\ \bibnamefont {Settineri}}, \bibinfo {author} {\bibfnamefont {Salvatore}\ \bibnamefont {Savasta}}, \ and\ \bibinfo {author} {\bibfnamefont {Franco}\ \bibnamefont {Nori}},\ }\bibfield  {title} {\enquote {\bibinfo {title} {{Resonant Raman scattering of single molecules under simultaneous strong cavity coupling and ultrastrong optomechanical coupling in plasmonic resonators: Phonon-dressed polaritons}},}\ }\href {\doibase 10.1103/PhysRevB.104.045431} {\bibfield  {journal} {\bibinfo  {journal} {Phys. Rev. B}\ }\textbf {\bibinfo {volume} {104}},\ \bibinfo {pages} {045431} (\bibinfo {year} {2021})}\BibitemShut {NoStop}%
\bibitem [{\citenamefont {Esteban}\ \emph {et~al.}(2022)\citenamefont {Esteban}, \citenamefont {Baumberg},\ and\ \citenamefont {Aizpurua}}]{Esteban2022}%
  \BibitemOpen
  \bibfield  {author} {\bibinfo {author} {\bibfnamefont {Ruben}\ \bibnamefont {Esteban}}, \bibinfo {author} {\bibfnamefont {Jeremy~J.}\ \bibnamefont {Baumberg}}, \ and\ \bibinfo {author} {\bibfnamefont {Javier}\ \bibnamefont {Aizpurua}},\ }\bibfield  {title} {\enquote {\bibinfo {title} {{Molecular Optomechanics Approach to Surface-Enhanced Raman Scattering}},}\ }\href {\doibase 10.1021/acs.accounts.1c00759} {\bibfield  {journal} {\bibinfo  {journal} {Acc. Chem. Res.}\ }\textbf {\bibinfo {volume} {55}},\ \bibinfo {pages} {1889} (\bibinfo {year} {2022})}\BibitemShut {NoStop}%
\bibitem [{\citenamefont {Lombardi}\ \emph {et~al.}(2018)\citenamefont {Lombardi}, \citenamefont {Schmidt}, \citenamefont {Weller}, \citenamefont {Deacon}, \citenamefont {Benz}, \citenamefont {de~Nijs}, \citenamefont {Aizpurua},\ and\ \citenamefont {Baumberg}}]{Anna-PRX}%
  \BibitemOpen
  \bibfield  {author} {\bibinfo {author} {\bibfnamefont {Anna}\ \bibnamefont {Lombardi}}, \bibinfo {author} {\bibfnamefont {Miko\l{}aj~K.}\ \bibnamefont {Schmidt}}, \bibinfo {author} {\bibfnamefont {Lee}\ \bibnamefont {Weller}}, \bibinfo {author} {\bibfnamefont {William~M.}\ \bibnamefont {Deacon}}, \bibinfo {author} {\bibfnamefont {Felix}\ \bibnamefont {Benz}}, \bibinfo {author} {\bibfnamefont {Bart}\ \bibnamefont {de~Nijs}}, \bibinfo {author} {\bibfnamefont {Javier}\ \bibnamefont {Aizpurua}}, \ and\ \bibinfo {author} {\bibfnamefont {Jeremy~J.}\ \bibnamefont {Baumberg}},\ }\bibfield  {title} {\enquote {\bibinfo {title} {Pulsed molecular optomechanics in plasmonic nanocavities: From nonlinear vibrational instabilities to bond-breaking},}\ }\href {\doibase 10.1103/PhysRevX.8.011016} {\bibfield  {journal} {\bibinfo  {journal} {Phys. Rev. X}\ }\textbf {\bibinfo {volume} {8}},\ \bibinfo {pages} {011016} (\bibinfo {year} {2018})}\BibitemShut {NoStop}%
\bibitem [{\citenamefont {Shishkov}\ \emph {et~al.}(2024{\natexlab{a}})\citenamefont {Shishkov}, \citenamefont {Andrianov},\ and\ \citenamefont {Zasedatelev}}]{Shishkov-Room}%
  \BibitemOpen
  \bibfield  {author} {\bibinfo {author} {\bibfnamefont {Vladislav~Yu.}\ \bibnamefont {Shishkov}}, \bibinfo {author} {\bibfnamefont {Evgeny~S.}\ \bibnamefont {Andrianov}}, \ and\ \bibinfo {author} {\bibfnamefont {Anton~V.}\ \bibnamefont {Zasedatelev}},\ }\bibfield  {title} {\enquote {\bibinfo {title} {Room-temperature optomechanics with light-matter condensates},}\ }\href {\doibase 10.1103/PhysRevB.110.134321} {\bibfield  {journal} {\bibinfo  {journal} {Phys. Rev. B}\ }\textbf {\bibinfo {volume} {110}},\ \bibinfo {pages} {134321} (\bibinfo {year} {2024}{\natexlab{a}})}\BibitemShut {NoStop}%
\bibitem [{\citenamefont {Shishkov}\ \emph {et~al.}(2024{\natexlab{b}})\citenamefont {Shishkov}, \citenamefont {Andrianov}, \citenamefont {Tretiak}, \citenamefont {Whaley},\ and\ \citenamefont {Zasedatelev}}]{Shishkov-PRL}%
  \BibitemOpen
  \bibfield  {author} {\bibinfo {author} {\bibfnamefont {Vladislav~Yu.}\ \bibnamefont {Shishkov}}, \bibinfo {author} {\bibfnamefont {Evgeny~S.}\ \bibnamefont {Andrianov}}, \bibinfo {author} {\bibfnamefont {Sergei}\ \bibnamefont {Tretiak}}, \bibinfo {author} {\bibfnamefont {K.~Birgitta}\ \bibnamefont {Whaley}}, \ and\ \bibinfo {author} {\bibfnamefont {Anton~V.}\ \bibnamefont {Zasedatelev}},\ }\bibfield  {title} {\enquote {\bibinfo {title} {{Sympathetic Mechanism for Vibrational Condensation Enabled by Polariton Optomechanical Interaction}},}\ }\href {\doibase 10.1103/PhysRevLett.133.186903} {\bibfield  {journal} {\bibinfo  {journal} {Phys. Rev. Lett.}\ }\textbf {\bibinfo {volume} {133}},\ \bibinfo {pages} {186903} (\bibinfo {year} {2024}{\natexlab{b}})}\BibitemShut {NoStop}%
\bibitem [{\citenamefont {Carlon~Zambon}\ \emph {et~al.}(2022)\citenamefont {Carlon~Zambon}, \citenamefont {Denis}, \citenamefont {De~Oliveira}, \citenamefont {Ravets}, \citenamefont {Ciuti}, \citenamefont {Favero},\ and\ \citenamefont {Bloch}}]{Carlon-PRL}%
  \BibitemOpen
  \bibfield  {author} {\bibinfo {author} {\bibfnamefont {N.}~\bibnamefont {Carlon~Zambon}}, \bibinfo {author} {\bibfnamefont {Z.}~\bibnamefont {Denis}}, \bibinfo {author} {\bibfnamefont {R.}~\bibnamefont {De~Oliveira}}, \bibinfo {author} {\bibfnamefont {S.}~\bibnamefont {Ravets}}, \bibinfo {author} {\bibfnamefont {C.}~\bibnamefont {Ciuti}}, \bibinfo {author} {\bibfnamefont {I.}~\bibnamefont {Favero}}, \ and\ \bibinfo {author} {\bibfnamefont {J.}~\bibnamefont {Bloch}},\ }\bibfield  {title} {\enquote {\bibinfo {title} {Enhanced cavity optomechanics with quantum-well exciton polaritons},}\ }\href {\doibase 10.1103/PhysRevLett.129.093603} {\bibfield  {journal} {\bibinfo  {journal} {Phys. Rev. Lett.}\ }\textbf {\bibinfo {volume} {129}},\ \bibinfo {pages} {093603} (\bibinfo {year} {2022})}\BibitemShut {NoStop}%
\bibitem [{\citenamefont {Jusserand}\ \emph {et~al.}(2015)\citenamefont {Jusserand}, \citenamefont {Poddubny}, \citenamefont {Poshakinskiy}, \citenamefont {Fainstein},\ and\ \citenamefont {Lemaitre}}]{Jusserand-PRL}%
  \BibitemOpen
  \bibfield  {author} {\bibinfo {author} {\bibfnamefont {B.}~\bibnamefont {Jusserand}}, \bibinfo {author} {\bibfnamefont {A.~N.}\ \bibnamefont {Poddubny}}, \bibinfo {author} {\bibfnamefont {A.~V.}\ \bibnamefont {Poshakinskiy}}, \bibinfo {author} {\bibfnamefont {A.}~\bibnamefont {Fainstein}}, \ and\ \bibinfo {author} {\bibfnamefont {A.}~\bibnamefont {Lemaitre}},\ }\bibfield  {title} {\enquote {\bibinfo {title} {Polariton resonances for ultrastrong coupling cavity optomechanics in $\mathrm{GaAs}/\mathrm{AlAs}$ multiple quantum wells},}\ }\href {\doibase 10.1103/PhysRevLett.115.267402} {\bibfield  {journal} {\bibinfo  {journal} {Phys. Rev. Lett.}\ }\textbf {\bibinfo {volume} {115}},\ \bibinfo {pages} {267402} (\bibinfo {year} {2015})}\BibitemShut {NoStop}%
\bibitem [{\citenamefont {Kyriienko}\ \emph {et~al.}(2014)\citenamefont {Kyriienko}, \citenamefont {Liew},\ and\ \citenamefont {Shelykh}}]{Kyriienko-PRL}%
  \BibitemOpen
  \bibfield  {author} {\bibinfo {author} {\bibfnamefont {O.}~\bibnamefont {Kyriienko}}, \bibinfo {author} {\bibfnamefont {T.~C.~H.}\ \bibnamefont {Liew}}, \ and\ \bibinfo {author} {\bibfnamefont {I.~A.}\ \bibnamefont {Shelykh}},\ }\bibfield  {title} {\enquote {\bibinfo {title} {Optomechanics with cavity polaritons: Dissipative coupling and unconventional bistability},}\ }\href {\doibase 10.1103/PhysRevLett.112.076402} {\bibfield  {journal} {\bibinfo  {journal} {Phys. Rev. Lett.}\ }\textbf {\bibinfo {volume} {112}},\ \bibinfo {pages} {076402} (\bibinfo {year} {2014})}\BibitemShut {NoStop}%
\bibitem [{\citenamefont {Huang}\ and\ \citenamefont {Zhang}(2025)}]{Jian-2025}%
  \BibitemOpen
  \bibfield  {author} {\bibinfo {author} {\bibfnamefont {Jian}\ \bibnamefont {Huang}}\ and\ \bibinfo {author} {\bibfnamefont {Zhedong}\ \bibnamefont {Zhang}},\ }\bibfield  {title} {\enquote {\bibinfo {title} {Room-temperature exciton-vibration-photon entanglement in polariton optomechanics},}\ }\href {\doibase 10.1103/2dfx-23qp} {\bibfield  {journal} {\bibinfo  {journal} {Phys. Rev. A}\ }\textbf {\bibinfo {volume} {112}},\ \bibinfo {pages} {013509} (\bibinfo {year} {2025})}\BibitemShut {NoStop}%
\bibitem [{\citenamefont {Picht}(1929)}]{Picht-1929}%
  \BibitemOpen
  \bibfield  {author} {\bibinfo {author} {\bibfnamefont {Johannes}\ \bibnamefont {Picht}},\ }\bibfield  {title} {\enquote {\bibinfo {title} {Beitrag zur theorie der totalreflexion},}\ }\href {\doibase https://doi.org/10.1002/andp.19293950402} {\bibfield  {journal} {\bibinfo  {journal} {Ann. Phys. (Leipzig)}\ }\textbf {\bibinfo {volume} {395}},\ \bibinfo {pages} {433} (\bibinfo {year} {1929})}\BibitemShut {NoStop}%
\bibitem [{\citenamefont {Goos}\ and\ \citenamefont {Hänchen}(1943)}]{Goos-1943}%
  \BibitemOpen
  \bibfield  {author} {\bibinfo {author} {\bibfnamefont {F.}~\bibnamefont {Goos}}\ and\ \bibinfo {author} {\bibfnamefont {H.}~\bibnamefont {Hänchen}},\ }\bibfield  {title} {\enquote {\bibinfo {title} {Über das eindringen des totalreflektierten lichtes in das dünnere medium},}\ }\href {\doibase https://doi.org/10.1002/andp.19434350504} {\bibfield  {journal} {\bibinfo  {journal} {Ann. Phys. (Leipzig)}\ }\textbf {\bibinfo {volume} {435}},\ \bibinfo {pages} {383} (\bibinfo {year} {1943})}\BibitemShut {NoStop}%
\bibitem [{\citenamefont {Goos}\ and\ \citenamefont {Hänchen}(1947)}]{Goos-1947}%
  \BibitemOpen
  \bibfield  {author} {\bibinfo {author} {\bibfnamefont {F.}~\bibnamefont {Goos}}\ and\ \bibinfo {author} {\bibfnamefont {H.}~\bibnamefont {Hänchen}},\ }\bibfield  {title} {\enquote {\bibinfo {title} {Ein neuer und fundamentaler versuch zur totalreflexion},}\ }\href {\doibase https://doi.org/10.1002/andp.19474360704} {\bibfield  {journal} {\bibinfo  {journal} {Ann. Phys. (Leipzig)}\ }\textbf {\bibinfo {volume} {436}},\ \bibinfo {pages} {333} (\bibinfo {year} {1947})}\BibitemShut {NoStop}%
\bibitem [{\citenamefont {Wang}\ \emph {et~al.}(2008{\natexlab{a}})\citenamefont {Wang}, \citenamefont {Cao}, \citenamefont {Li}, \citenamefont {Hao}, \citenamefont {Yu},\ and\ \citenamefont {Shen}}]{Yi-2008}%
  \BibitemOpen
  \bibfield  {author} {\bibinfo {author} {\bibfnamefont {Yi}~\bibnamefont {Wang}}, \bibinfo {author} {\bibfnamefont {Zhuangqi}\ \bibnamefont {Cao}}, \bibinfo {author} {\bibfnamefont {Honggen}\ \bibnamefont {Li}}, \bibinfo {author} {\bibfnamefont {Jun}\ \bibnamefont {Hao}}, \bibinfo {author} {\bibfnamefont {Tianyi}\ \bibnamefont {Yu}}, \ and\ \bibinfo {author} {\bibfnamefont {Qishun}\ \bibnamefont {Shen}},\ }\bibfield  {title} {\enquote {\bibinfo {title} {Electric control of spatial beam position based on the {Goos–Hänchen} effect},}\ }\href {\doibase 10.1063/1.2977873} {\bibfield  {journal} {\bibinfo  {journal} {Appl. Phys. Lett.}\ }\textbf {\bibinfo {volume} {93}},\ \bibinfo {pages} {091103} (\bibinfo {year} {2008}{\natexlab{a}})}\BibitemShut {NoStop}%
\bibitem [{\citenamefont {Hashimoto}\ and\ \citenamefont {Yoshino}(1989)}]{Hashimoto:89}%
  \BibitemOpen
  \bibfield  {author} {\bibinfo {author} {\bibfnamefont {Takaharu}\ \bibnamefont {Hashimoto}}\ and\ \bibinfo {author} {\bibfnamefont {Toshihiko}\ \bibnamefont {Yoshino}},\ }\bibfield  {title} {\enquote {\bibinfo {title} {Optical heterodyne sensor using the {Goos--H\"{a}nchen} shift},}\ }\href {\doibase 10.1364/OL.14.000913} {\bibfield  {journal} {\bibinfo  {journal} {Opt. Lett.}\ }\textbf {\bibinfo {volume} {14}},\ \bibinfo {pages} {913} (\bibinfo {year} {1989})}\BibitemShut {NoStop}%
\bibitem [{\citenamefont {Chen}\ \emph {et~al.}(2007)\citenamefont {Chen}, \citenamefont {Lin}, \citenamefont {Liao}, \citenamefont {Lin}, \citenamefont {Chiang}, \citenamefont {Leung}, \citenamefont {Sijercic},\ and\ \citenamefont {Tse}}]{Chen-07}%
  \BibitemOpen
  \bibfield  {author} {\bibinfo {author} {\bibfnamefont {Chih-Wei}\ \bibnamefont {Chen}}, \bibinfo {author} {\bibfnamefont {Wen-Chi}\ \bibnamefont {Lin}}, \bibinfo {author} {\bibfnamefont {Lu-Shing}\ \bibnamefont {Liao}}, \bibinfo {author} {\bibfnamefont {Zheng-Hung}\ \bibnamefont {Lin}}, \bibinfo {author} {\bibfnamefont {Hai-Pang}\ \bibnamefont {Chiang}}, \bibinfo {author} {\bibfnamefont {Pui-Tak}\ \bibnamefont {Leung}}, \bibinfo {author} {\bibfnamefont {Edin}\ \bibnamefont {Sijercic}}, \ and\ \bibinfo {author} {\bibfnamefont {Wan-Sun}\ \bibnamefont {Tse}},\ }\bibfield  {title} {\enquote {\bibinfo {title} {Optical temperature sensing based on the {Goos-H\"{a}nchen} effect},}\ }\href {\doibase 10.1364/AO.46.005347} {\bibfield  {journal} {\bibinfo  {journal} {Appl. Opt.}\ }\textbf {\bibinfo {volume} {46}},\ \bibinfo {pages} {5347} (\bibinfo {year} {2007})}\BibitemShut {NoStop}%
\bibitem [{\citenamefont {Yin}\ and\ \citenamefont {Hesselink}(2006)}]{Yin-2006}%
  \BibitemOpen
  \bibfield  {author} {\bibinfo {author} {\bibfnamefont {Xiaobo}\ \bibnamefont {Yin}}\ and\ \bibinfo {author} {\bibfnamefont {Lambertus}\ \bibnamefont {Hesselink}},\ }\bibfield  {title} {\enquote {\bibinfo {title} {{Goos-H\"anchen} shift surface plasmon resonance sensor},}\ }\href {\doibase 10.1063/1.2424277} {\bibfield  {journal} {\bibinfo  {journal} {Appl. Phys. Lett.}\ }\textbf {\bibinfo {volume} {89}},\ \bibinfo {pages} {261108} (\bibinfo {year} {2006})}\BibitemShut {NoStop}%
\bibitem [{\citenamefont {Renard}(1964)}]{Renard:64}%
  \BibitemOpen
  \bibfield  {author} {\bibinfo {author} {\bibfnamefont {R\'{e}mi~H.}\ \bibnamefont {Renard}},\ }\bibfield  {title} {\enquote {\bibinfo {title} {Total reflection: A new evaluation of the {Goos--H\"{a}nchen} shift},}\ }\href {\doibase 10.1364/JOSA.54.001190} {\bibfield  {journal} {\bibinfo  {journal} {J. Opt. Soc. Am.}\ }\textbf {\bibinfo {volume} {54}},\ \bibinfo {pages} {1190} (\bibinfo {year} {1964})}\BibitemShut {NoStop}%
\bibitem [{\citenamefont {Qing}\ and\ \citenamefont {Chen}(2004)}]{Qing:04}%
  \BibitemOpen
  \bibfield  {author} {\bibinfo {author} {\bibfnamefont {De-Kui}\ \bibnamefont {Qing}}\ and\ \bibinfo {author} {\bibfnamefont {Gang}\ \bibnamefont {Chen}},\ }\bibfield  {title} {\enquote {\bibinfo {title} {{Goos-H\"{a}nchen} shifts at the interfaces between left- and right-handed media},}\ }\href {\doibase 10.1364/OL.29.000872} {\bibfield  {journal} {\bibinfo  {journal} {Opt. Lett.}\ }\textbf {\bibinfo {volume} {29}},\ \bibinfo {pages} {872} (\bibinfo {year} {2004})}\BibitemShut {NoStop}%
\bibitem [{\citenamefont {Tamir}\ and\ \citenamefont {Bertoni}(1971)}]{Tamir:71}%
  \BibitemOpen
  \bibfield  {author} {\bibinfo {author} {\bibfnamefont {T.}~\bibnamefont {Tamir}}\ and\ \bibinfo {author} {\bibfnamefont {H.~L.}\ \bibnamefont {Bertoni}},\ }\bibfield  {title} {\enquote {\bibinfo {title} {Lateral displacement of optical beams at multilayered and periodic structures$\ast$},}\ }\href {\doibase 10.1364/JOSA.61.001397} {\bibfield  {journal} {\bibinfo  {journal} {J. Opt. Soc. Am.}\ }\textbf {\bibinfo {volume} {61}},\ \bibinfo {pages} {1397} (\bibinfo {year} {1971})}\BibitemShut {NoStop}%
\bibitem [{\citenamefont {Chu}\ and\ \citenamefont {Wong}(1982)}]{Chu-82}%
  \BibitemOpen
  \bibfield  {author} {\bibinfo {author} {\bibfnamefont {S.}~\bibnamefont {Chu}}\ and\ \bibinfo {author} {\bibfnamefont {S.}~\bibnamefont {Wong}},\ }\bibfield  {title} {\enquote {\bibinfo {title} {Linear pulse propagation in an absorbing medium},}\ }\href {\doibase 10.1103/PhysRevLett.48.738} {\bibfield  {journal} {\bibinfo  {journal} {Phys. Rev. Lett.}\ }\textbf {\bibinfo {volume} {48}},\ \bibinfo {pages} {738} (\bibinfo {year} {1982})}\BibitemShut {NoStop}%
\bibitem [{\citenamefont {Birman}\ \emph {et~al.}(1983)\citenamefont {Birman}, \citenamefont {Pattanayak},\ and\ \citenamefont {Puri}}]{Birman-83}%
  \BibitemOpen
  \bibfield  {author} {\bibinfo {author} {\bibfnamefont {Joseph~L.}\ \bibnamefont {Birman}}, \bibinfo {author} {\bibfnamefont {Deva~N.}\ \bibnamefont {Pattanayak}}, \ and\ \bibinfo {author} {\bibfnamefont {Ashok}\ \bibnamefont {Puri}},\ }\bibfield  {title} {\enquote {\bibinfo {title} {Prediction of a resonance-enhanced laser-beam displacement at total internal reflection in semiconductors},}\ }\href {\doibase 10.1103/PhysRevLett.50.1664} {\bibfield  {journal} {\bibinfo  {journal} {Phys. Rev. Lett.}\ }\textbf {\bibinfo {volume} {50}},\ \bibinfo {pages} {1664} (\bibinfo {year} {1983})}\BibitemShut {NoStop}%
\bibitem [{\citenamefont {Berman}(2002)}]{Berman-02}%
  \BibitemOpen
  \bibfield  {author} {\bibinfo {author} {\bibfnamefont {P.~R.}\ \bibnamefont {Berman}},\ }\bibfield  {title} {\enquote {\bibinfo {title} {{Goos-H\"anchen} shift in negatively refractive media},}\ }\href {\doibase 10.1103/PhysRevE.66.067603} {\bibfield  {journal} {\bibinfo  {journal} {Phys. Rev. E}\ }\textbf {\bibinfo {volume} {66}},\ \bibinfo {pages} {067603} (\bibinfo {year} {2002})}\BibitemShut {NoStop}%
\bibitem [{\citenamefont {Pfleghaar}\ \emph {et~al.}(1993)\citenamefont {Pfleghaar}, \citenamefont {Marseille},\ and\ \citenamefont {Weis}}]{Pfleghaar-93}%
  \BibitemOpen
  \bibfield  {author} {\bibinfo {author} {\bibfnamefont {E.}~\bibnamefont {Pfleghaar}}, \bibinfo {author} {\bibfnamefont {A.}~\bibnamefont {Marseille}}, \ and\ \bibinfo {author} {\bibfnamefont {A.}~\bibnamefont {Weis}},\ }\bibfield  {title} {\enquote {\bibinfo {title} {Quantitative investigation of the effect of resonant absorbers on the {Goos-H\"anchen} shift},}\ }\href {\doibase 10.1103/PhysRevLett.70.2281} {\bibfield  {journal} {\bibinfo  {journal} {Phys. Rev. Lett.}\ }\textbf {\bibinfo {volume} {70}},\ \bibinfo {pages} {2281} (\bibinfo {year} {1993})}\BibitemShut {NoStop}%
\bibitem [{\citenamefont {Soboleva}\ \emph {et~al.}(2012)\citenamefont {Soboleva}, \citenamefont {Moskalenko},\ and\ \citenamefont {Fedyanin}}]{Soboleva-2012}%
  \BibitemOpen
  \bibfield  {author} {\bibinfo {author} {\bibfnamefont {I.~V.}\ \bibnamefont {Soboleva}}, \bibinfo {author} {\bibfnamefont {V.~V.}\ \bibnamefont {Moskalenko}}, \ and\ \bibinfo {author} {\bibfnamefont {A.~A.}\ \bibnamefont {Fedyanin}},\ }\bibfield  {title} {\enquote {\bibinfo {title} {Giant {Goos-H\"anchen} effect and fano resonance at photonic crystal surfaces},}\ }\href {\doibase 10.1103/PhysRevLett.108.123901} {\bibfield  {journal} {\bibinfo  {journal} {Phys. Rev. Lett.}\ }\textbf {\bibinfo {volume} {108}},\ \bibinfo {pages} {123901} (\bibinfo {year} {2012})}\BibitemShut {NoStop}%
\bibitem [{\citenamefont {Wan}\ \emph {et~al.}(2012)\citenamefont {Wan}, \citenamefont {Zheng}, \citenamefont {Kong}, \citenamefont {Zhao}, \citenamefont {Liu}, \citenamefont {Bian},\ and\ \citenamefont {Liu}}]{Wan:12}%
  \BibitemOpen
  \bibfield  {author} {\bibinfo {author} {\bibfnamefont {Yuhang}\ \bibnamefont {Wan}}, \bibinfo {author} {\bibfnamefont {Zheng}\ \bibnamefont {Zheng}}, \bibinfo {author} {\bibfnamefont {Weijing}\ \bibnamefont {Kong}}, \bibinfo {author} {\bibfnamefont {Xin}\ \bibnamefont {Zhao}}, \bibinfo {author} {\bibfnamefont {Ya}~\bibnamefont {Liu}}, \bibinfo {author} {\bibfnamefont {Yusheng}\ \bibnamefont {Bian}}, \ and\ \bibinfo {author} {\bibfnamefont {Jiansheng}\ \bibnamefont {Liu}},\ }\bibfield  {title} {\enquote {\bibinfo {title} {Nearly three orders of magnitude enhancement of {Goos-H\"anchen} shift by exciting bloch surface wave},}\ }\href {\doibase 10.1364/OE.20.008998} {\bibfield  {journal} {\bibinfo  {journal} {Opt. Express}\ }\textbf {\bibinfo {volume} {20}},\ \bibinfo {pages} {8998} (\bibinfo {year} {2012})}\BibitemShut {NoStop}%
\bibitem [{\citenamefont {Yang}\ \emph {et~al.}(2014)\citenamefont {Yang}, \citenamefont {Zhu},\ and\ \citenamefont {Li}}]{Yang:14}%
  \BibitemOpen
  \bibfield  {author} {\bibinfo {author} {\bibfnamefont {Rui}\ \bibnamefont {Yang}}, \bibinfo {author} {\bibfnamefont {Wenkan}\ \bibnamefont {Zhu}}, \ and\ \bibinfo {author} {\bibfnamefont {Jingjing}\ \bibnamefont {Li}},\ }\bibfield  {title} {\enquote {\bibinfo {title} {Giant positive and negative {Goos-H\"{a}nchen} shift on dielectric gratings caused by guided mode resonance},}\ }\href {\doibase 10.1364/OE.22.002043} {\bibfield  {journal} {\bibinfo  {journal} {Opt. Express}\ }\textbf {\bibinfo {volume} {22}},\ \bibinfo {pages} {2043} (\bibinfo {year} {2014})}\BibitemShut {NoStop}%
\bibitem [{\citenamefont {Bonnet}\ \emph {et~al.}(2001)\citenamefont {Bonnet}, \citenamefont {Chauvat}, \citenamefont {Emile}, \citenamefont {Bretenaker}, \citenamefont {Floch},\ and\ \citenamefont {Dutriaux}}]{Bonnet:01}%
  \BibitemOpen
  \bibfield  {author} {\bibinfo {author} {\bibfnamefont {Christophe}\ \bibnamefont {Bonnet}}, \bibinfo {author} {\bibfnamefont {Dominique}\ \bibnamefont {Chauvat}}, \bibinfo {author} {\bibfnamefont {Olivier}\ \bibnamefont {Emile}}, \bibinfo {author} {\bibfnamefont {Fabien}\ \bibnamefont {Bretenaker}}, \bibinfo {author} {\bibfnamefont {Albert~Le}\ \bibnamefont {Floch}}, \ and\ \bibinfo {author} {\bibfnamefont {Laurent}\ \bibnamefont {Dutriaux}},\ }\bibfield  {title} {\enquote {\bibinfo {title} {{Measurement of positive and negative Goos--H\"{a}nchen effects for metallic gratings near Wood anomalies}},}\ }\href {\doibase 10.1364/OL.26.000666} {\bibfield  {journal} {\bibinfo  {journal} {Opt. Lett.}\ }\textbf {\bibinfo {volume} {26}},\ \bibinfo {pages} {666} (\bibinfo {year} {2001})}\BibitemShut {NoStop}%
\bibitem [{\citenamefont {Yin}\ \emph {et~al.}(2004)\citenamefont {Yin}, \citenamefont {Hesselink}, \citenamefont {Liu}, \citenamefont {Fang},\ and\ \citenamefont {Zhang}}]{Yin-2004}%
  \BibitemOpen
  \bibfield  {author} {\bibinfo {author} {\bibfnamefont {Xiaobo}\ \bibnamefont {Yin}}, \bibinfo {author} {\bibfnamefont {Lambertus}\ \bibnamefont {Hesselink}}, \bibinfo {author} {\bibfnamefont {Zhaowei}\ \bibnamefont {Liu}}, \bibinfo {author} {\bibfnamefont {Nicholas}\ \bibnamefont {Fang}}, \ and\ \bibinfo {author} {\bibfnamefont {Xiang}\ \bibnamefont {Zhang}},\ }\bibfield  {title} {\enquote {\bibinfo {title} {Large positive and negative lateral optical beam displacements due to surface plasmon resonance},}\ }\href {\doibase 10.1063/1.1775294} {\bibfield  {journal} {\bibinfo  {journal} {Appl. Phys. Lett.}\ }\textbf {\bibinfo {volume} {85}},\ \bibinfo {pages} {372} (\bibinfo {year} {2004})}\BibitemShut {NoStop}%
\bibitem [{\citenamefont {Xu}\ \emph {et~al.}(2015)\citenamefont {Xu}, \citenamefont {Chan},\ and\ \citenamefont {Chen}}]{Xu-2015}%
  \BibitemOpen
  \bibfield  {author} {\bibinfo {author} {\bibfnamefont {Yadong}\ \bibnamefont {Xu}}, \bibinfo {author} {\bibfnamefont {C.~T.}\ \bibnamefont {Chan}}, \ and\ \bibinfo {author} {\bibfnamefont {Huanyang}\ \bibnamefont {Chen}},\ }\bibfield  {title} {\enquote {\bibinfo {title} {{Goos-H{\"a}nchen} effect in epsilon-near-zero metamaterials},}\ }\href {\doibase 10.1038/srep08681} {\bibfield  {journal} {\bibinfo  {journal} {Sci. Rep.}\ }\textbf {\bibinfo {volume} {5}},\ \bibinfo {pages} {8681} (\bibinfo {year} {2015})}\BibitemShut {NoStop}%
\bibitem [{\citenamefont {Chen}\ \emph {et~al.}(2016)\citenamefont {Chen}, \citenamefont {Bian}, \citenamefont {Chiang},\ and\ \citenamefont {Leung}}]{Chen_2016}%
  \BibitemOpen
  \bibfield  {author} {\bibinfo {author} {\bibfnamefont {Chih-Wei}\ \bibnamefont {Chen}}, \bibinfo {author} {\bibfnamefont {Tingting}\ \bibnamefont {Bian}}, \bibinfo {author} {\bibfnamefont {Hai-Pang}\ \bibnamefont {Chiang}}, \ and\ \bibinfo {author} {\bibfnamefont {P~T}\ \bibnamefont {Leung}},\ }\bibfield  {title} {\enquote {\bibinfo {title} {Nonlocal optical effects on the {Goos-H\"anchen} shifts at multilayered hyperbolic metamaterials},}\ }\href {\doibase 10.1088/2040-8978/18/2/025104} {\bibfield  {journal} {\bibinfo  {journal} {J. Opt.}\ }\textbf {\bibinfo {volume} {18}},\ \bibinfo {pages} {025104} (\bibinfo {year} {2016})}\BibitemShut {NoStop}%
\bibitem [{\citenamefont {Idrees}\ \emph {et~al.}(2023)\citenamefont {Idrees}, \citenamefont {Ullah},\ and\ \citenamefont {Wang}}]{Idress-PRA}%
  \BibitemOpen
  \bibfield  {author} {\bibinfo {author} {\bibfnamefont {Muhammad}\ \bibnamefont {Idrees}}, \bibinfo {author} {\bibfnamefont {Muhib}\ \bibnamefont {Ullah}}, \ and\ \bibinfo {author} {\bibfnamefont {Li-Gang}\ \bibnamefont {Wang}},\ }\bibfield  {title} {\enquote {\bibinfo {title} {Enhancement of the {Goos-H\"anchen} shift via chiral quantum-dot molecule systems},}\ }\href {\doibase 10.1103/PhysRevA.108.013701} {\bibfield  {journal} {\bibinfo  {journal} {Phys. Rev. A}\ }\textbf {\bibinfo {volume} {108}},\ \bibinfo {pages} {013701} (\bibinfo {year} {2023})}\BibitemShut {NoStop}%
\bibitem [{\citenamefont {Ziauddin}\ \emph {et~al.}(2010)\citenamefont {Ziauddin}, \citenamefont {Qamar},\ and\ \citenamefont {Zubairy}}]{Zia_2010_Coherent}%
  \BibitemOpen
  \bibfield  {author} {\bibinfo {author} {\bibnamefont {Ziauddin}}, \bibinfo {author} {\bibfnamefont {Sajid}\ \bibnamefont {Qamar}}, \ and\ \bibinfo {author} {\bibfnamefont {M.~Suhail}\ \bibnamefont {Zubairy}},\ }\bibfield  {title} {\enquote {\bibinfo {title} {{Coherent control of the Goos-H\"anchen shift}},}\ }\href {\doibase 10.1103/PhysRevA.81.023821} {\bibfield  {journal} {\bibinfo  {journal} {Phys. Rev. A}\ }\textbf {\bibinfo {volume} {81}},\ \bibinfo {pages} {023821} (\bibinfo {year} {2010})}\BibitemShut {NoStop}%
\bibitem [{\citenamefont {Ullah}\ \emph {et~al.}(2019)\citenamefont {Ullah}, \citenamefont {Abbas}, \citenamefont {Jing},\ and\ \citenamefont {Wang}}]{Muhib-PRA}%
  \BibitemOpen
  \bibfield  {author} {\bibinfo {author} {\bibfnamefont {Muhib}\ \bibnamefont {Ullah}}, \bibinfo {author} {\bibfnamefont {Adeel}\ \bibnamefont {Abbas}}, \bibinfo {author} {\bibfnamefont {Jun}\ \bibnamefont {Jing}}, \ and\ \bibinfo {author} {\bibfnamefont {Li-Gang}\ \bibnamefont {Wang}},\ }\bibfield  {title} {\enquote {\bibinfo {title} {Flexible manipulation of the {Goos-H\"anchen} shift in a cavity optomechanical system},}\ }\href {\doibase 10.1103/PhysRevA.100.063833} {\bibfield  {journal} {\bibinfo  {journal} {Phys. Rev. A}\ }\textbf {\bibinfo {volume} {100}},\ \bibinfo {pages} {063833} (\bibinfo {year} {2019})}\BibitemShut {NoStop}%
\bibitem [{\citenamefont {Khan}\ \emph {et~al.}(2020)\citenamefont {Khan}, \citenamefont {Abbas}, \citenamefont {Chaung}, \citenamefont {Ahmed},\ and\ \citenamefont {Ziauddin}}]{Anwar-PRA}%
  \BibitemOpen
  \bibfield  {author} {\bibinfo {author} {\bibfnamefont {Anwar~Ali}\ \bibnamefont {Khan}}, \bibinfo {author} {\bibfnamefont {Muqaddar}\ \bibnamefont {Abbas}}, \bibinfo {author} {\bibfnamefont {You-Lin}\ \bibnamefont {Chaung}}, \bibinfo {author} {\bibfnamefont {Iftikhar}\ \bibnamefont {Ahmed}}, \ and\ \bibinfo {author} {\bibnamefont {Ziauddin}},\ }\bibfield  {title} {\enquote {\bibinfo {title} {Investigation of the {Goos-H\"anchen} shift in an optomechanical cavity via quantum control},}\ }\href {\doibase 10.1103/PhysRevA.102.053718} {\bibfield  {journal} {\bibinfo  {journal} {Phys. Rev. A}\ }\textbf {\bibinfo {volume} {102}},\ \bibinfo {pages} {053718} (\bibinfo {year} {2020})}\BibitemShut {NoStop}%
\bibitem [{\citenamefont {Waseem}\ \emph {et~al.}(2024)\citenamefont {Waseem}, \citenamefont {Irfan},\ and\ \citenamefont {Qamar}}]{waseem_Goos_2024}%
  \BibitemOpen
  \bibfield  {author} {\bibinfo {author} {\bibfnamefont {Muhammad}\ \bibnamefont {Waseem}}, \bibinfo {author} {\bibfnamefont {Muhammad}\ \bibnamefont {Irfan}}, \ and\ \bibinfo {author} {\bibfnamefont {Shahid}\ \bibnamefont {Qamar}},\ }\bibfield  {title} {\enquote {\bibinfo {title} {{Magnomechanically controlled Goos-H\"anchen shift in cavity QED}},}\ }\href {\doibase 10.1103/PhysRevA.110.033711} {\bibfield  {journal} {\bibinfo  {journal} {Phys. Rev. A}\ }\textbf {\bibinfo {volume} {110}},\ \bibinfo {pages} {033711} (\bibinfo {year} {2024})}\BibitemShut {NoStop}%
\bibitem [{\citenamefont {Fahad}\ and\ \citenamefont {Xianlong}(2025)}]{fahad2025}%
  \BibitemOpen
  \bibfield  {author} {\bibinfo {author} {\bibfnamefont {Shah}\ \bibnamefont {Fahad}}\ and\ \bibinfo {author} {\bibfnamefont {Gao}\ \bibnamefont {Xianlong}},\ }\href {https://arxiv.org/abs/2511.20262} {\enquote {\bibinfo {title} {$\mathcal{PT}$-assisted control of {Goos-H\"anchen} shift in cavity magnomechanics},}\ } (\bibinfo {year} {2025}),\ \Eprint {http://arxiv.org/abs/2511.20262} {arXiv:2511.20262} \BibitemShut {NoStop}%
\bibitem [{\citenamefont {Reitz}\ \emph {et~al.}(2020)\citenamefont {Reitz}, \citenamefont {Sommer}, \citenamefont {Gurlek}, \citenamefont {Sandoghdar}, \citenamefont {Martin-Cano},\ and\ \citenamefont {Genes}}]{Reitz-2020}%
  \BibitemOpen
  \bibfield  {author} {\bibinfo {author} {\bibfnamefont {M.}~\bibnamefont {Reitz}}, \bibinfo {author} {\bibfnamefont {C.}~\bibnamefont {Sommer}}, \bibinfo {author} {\bibfnamefont {B.}~\bibnamefont {Gurlek}}, \bibinfo {author} {\bibfnamefont {V.}~\bibnamefont {Sandoghdar}}, \bibinfo {author} {\bibfnamefont {D.}~\bibnamefont {Martin-Cano}}, \ and\ \bibinfo {author} {\bibfnamefont {C.}~\bibnamefont {Genes}},\ }\bibfield  {title} {\enquote {\bibinfo {title} {Molecule-photon interactions in phononic environments},}\ }\href {\doibase 10.1103/PhysRevResearch.2.033270} {\bibfield  {journal} {\bibinfo  {journal} {Phys. Rev. Res.}\ }\textbf {\bibinfo {volume} {2}},\ \bibinfo {pages} {033270} (\bibinfo {year} {2020})}\BibitemShut {NoStop}%
\bibitem [{\citenamefont {Mart\'{\i}nez-Garc\'{\i}a}\ and\ \citenamefont {Mart\'{\i}n-Cano}(2024)}]{Mart-PRL}%
  \BibitemOpen
  \bibfield  {author} {\bibinfo {author} {\bibfnamefont {Miguel~\'A.}\ \bibnamefont {Mart\'{\i}nez-Garc\'{\i}a}}\ and\ \bibinfo {author} {\bibfnamefont {Diego}\ \bibnamefont {Mart\'{\i}n-Cano}},\ }\bibfield  {title} {\enquote {\bibinfo {title} {{Coherent Electron-Vibron Interactions in Surface-Enhanced Raman Scattering {(SERS)}}},}\ }\href {\doibase 10.1103/PhysRevLett.132.093601} {\bibfield  {journal} {\bibinfo  {journal} {Phys. Rev. Lett.}\ }\textbf {\bibinfo {volume} {132}},\ \bibinfo {pages} {093601} (\bibinfo {year} {2024})}\BibitemShut {NoStop}%
\bibitem [{\citenamefont {Spano}(2010)}]{Spano2010}%
  \BibitemOpen
  \bibfield  {author} {\bibinfo {author} {\bibfnamefont {Frank~C.}\ \bibnamefont {Spano}},\ }\bibfield  {title} {\enquote {\bibinfo {title} {The spectral signatures of frenkel polarons in h- and j-aggregates},}\ }\href {\doibase 10.1021/ar900233v} {\bibfield  {journal} {\bibinfo  {journal} {Acc. Chem. Res.}\ }\textbf {\bibinfo {volume} {43}},\ \bibinfo {pages} {429} (\bibinfo {year} {2010})}\BibitemShut {NoStop}%
\bibitem [{\citenamefont {Kurucz}\ and\ \citenamefont {M\o{}lmer}(2010)}]{Kurucz-PRA}%
  \BibitemOpen
  \bibfield  {author} {\bibinfo {author} {\bibfnamefont {Z.}~\bibnamefont {Kurucz}}\ and\ \bibinfo {author} {\bibfnamefont {K.}~\bibnamefont {M\o{}lmer}},\ }\bibfield  {title} {\enquote {\bibinfo {title} {Multilevel holstein-primakoff approximation and its application to atomic spin squeezing and ensemble quantum memories},}\ }\href {\doibase 10.1103/PhysRevA.81.032314} {\bibfield  {journal} {\bibinfo  {journal} {Phys. Rev. A}\ }\textbf {\bibinfo {volume} {81}},\ \bibinfo {pages} {032314} (\bibinfo {year} {2010})}\BibitemShut {NoStop}%
\bibitem [{\citenamefont {Huang}\ \emph {et~al.}(2023)\citenamefont {Huang}, \citenamefont {Liu}, \citenamefont {Xu},\ and\ \citenamefont {Liao}}]{huang2023}%
  \BibitemOpen
  \bibfield  {author} {\bibinfo {author} {\bibfnamefont {Jian}\ \bibnamefont {Huang}}, \bibinfo {author} {\bibfnamefont {Cheng}\ \bibnamefont {Liu}}, \bibinfo {author} {\bibfnamefont {Xun-Wei}\ \bibnamefont {Xu}}, \ and\ \bibinfo {author} {\bibfnamefont {Jie-Qiao}\ \bibnamefont {Liao}},\ }\href {https://arxiv.org/abs/2312.06274} {\enquote {\bibinfo {title} {Dark-mode theorems for quantum networks},}\ } (\bibinfo {year} {2023}),\ \Eprint {http://arxiv.org/abs/2312.06274} {arXiv:2312.06274 [quant-ph]} \BibitemShut {NoStop}%
\bibitem [{\citenamefont {Xiong}\ \emph {et~al.}(2015)\citenamefont {Xiong}, \citenamefont {Si}, \citenamefont {Lv}, \citenamefont {Yang},\ and\ \citenamefont {Wu}}]{Xiong2015}%
  \BibitemOpen
  \bibfield  {author} {\bibinfo {author} {\bibfnamefont {Hao}\ \bibnamefont {Xiong}}, \bibinfo {author} {\bibfnamefont {LiuGang}\ \bibnamefont {Si}}, \bibinfo {author} {\bibfnamefont {XinYou}\ \bibnamefont {Lv}}, \bibinfo {author} {\bibfnamefont {XiaoXue}\ \bibnamefont {Yang}}, \ and\ \bibinfo {author} {\bibfnamefont {Ying}\ \bibnamefont {Wu}},\ }\bibfield  {title} {\enquote {\bibinfo {title} {{Review of cavity optomechanics in the weak-coupling regime: from linearization to intrinsic nonlinear interactions}},}\ }\href {\doibase 10.1007/s11433-015-5648-9} {\bibfield  {journal} {\bibinfo  {journal} {Sci. China: Phys. Mech. Astron.}\ }\textbf {\bibinfo {volume} {58}},\ \bibinfo {pages} {1} (\bibinfo {year} {2015})}\BibitemShut {NoStop}%
\bibitem [{\citenamefont {Chen}(2023)}]{Chen_2023}%
  \BibitemOpen
  \bibfield  {author} {\bibinfo {author} {\bibfnamefont {Fujun}\ \bibnamefont {Chen}},\ }\bibfield  {title} {\enquote {\bibinfo {title} {{Electromagnetically induced grating in a nonlinear optomechanical cavity}},}\ }\href {\doibase 10.1088/1612-202X/ace70d} {\bibfield  {journal} {\bibinfo  {journal} {Laser Phys. Lett.}\ }\textbf {\bibinfo {volume} {20}},\ \bibinfo {pages} {095206} (\bibinfo {year} {2023})}\BibitemShut {NoStop}%
\bibitem [{\citenamefont {Li}\ \emph {et~al.}(2016)\citenamefont {Li}, \citenamefont {Nie},\ and\ \citenamefont {Chen}}]{Li_2016Transparency}%
  \BibitemOpen
  \bibfield  {author} {\bibinfo {author} {\bibfnamefont {Ling}\ \bibnamefont {Li}}, \bibinfo {author} {\bibfnamefont {Wenjie}\ \bibnamefont {Nie}}, \ and\ \bibinfo {author} {\bibfnamefont {Aixi}\ \bibnamefont {Chen}},\ }\bibfield  {title} {\enquote {\bibinfo {title} {{Transparency and tunable slow and fast light in a nonlinear optomechanical cavity}},}\ }\href {\doibase 10.1038/srep35090} {\bibfield  {journal} {\bibinfo  {journal} {Sci. Rep.}\ }\textbf {\bibinfo {volume} {6}},\ \bibinfo {pages} {35090} (\bibinfo {year} {2016})}\BibitemShut {NoStop}%
\bibitem [{\citenamefont {Walls}\ and\ \citenamefont {Milburn}(1994)}]{walls1994quantum}%
  \BibitemOpen
  \bibfield  {author} {\bibinfo {author} {\bibfnamefont {D.~F.}\ \bibnamefont {Walls}}\ and\ \bibinfo {author} {\bibfnamefont {G.~J.}\ \bibnamefont {Milburn}},\ }\href {\doibase 10.1007/978-3-642-79504-6} {\emph {\bibinfo {title} {{Quantum Optics}}}}\ (\bibinfo  {publisher} {Springer-Verlag},\ \bibinfo {address} {Berlin, Heidelberg},\ \bibinfo {year} {1994})\BibitemShut {NoStop}%
\bibitem [{\citenamefont {Artmann}(1948)}]{Artmann-1948}%
  \BibitemOpen
  \bibfield  {author} {\bibinfo {author} {\bibfnamefont {Kurt}\ \bibnamefont {Artmann}},\ }\bibfield  {title} {\enquote {\bibinfo {title} {Berechnung der seitenversetzung des totalreflektierten strahles},}\ }\href {\doibase https://doi.org/10.1002/andp.19484370108} {\bibfield  {journal} {\bibinfo  {journal} {Ann. Phys. (Leipzig)}\ }\textbf {\bibinfo {volume} {437}},\ \bibinfo {pages} {87} (\bibinfo {year} {1948})}\BibitemShut {NoStop}%
\bibitem [{\citenamefont {Li}(2003)}]{Li-2003PRL}%
  \BibitemOpen
  \bibfield  {author} {\bibinfo {author} {\bibfnamefont {Chun-Fang}\ \bibnamefont {Li}},\ }\bibfield  {title} {\enquote {\bibinfo {title} {Negative lateral shift of a light beam transmitted through a dielectric slab and interaction of boundary effects},}\ }\href {\doibase 10.1103/PhysRevLett.91.133903} {\bibfield  {journal} {\bibinfo  {journal} {Phys. Rev. Lett.}\ }\textbf {\bibinfo {volume} {91}},\ \bibinfo {pages} {133903} (\bibinfo {year} {2003})}\BibitemShut {NoStop}%
\bibitem [{\citenamefont {Wang}\ \emph {et~al.}(2005)\citenamefont {Wang}, \citenamefont {Chen},\ and\ \citenamefont {Zhu}}]{Wang:05}%
  \BibitemOpen
  \bibfield  {author} {\bibinfo {author} {\bibfnamefont {Li-Gang}\ \bibnamefont {Wang}}, \bibinfo {author} {\bibfnamefont {Hong}\ \bibnamefont {Chen}}, \ and\ \bibinfo {author} {\bibfnamefont {Shi-Yao}\ \bibnamefont {Zhu}},\ }\bibfield  {title} {\enquote {\bibinfo {title} {Large negative {Goos--H\"{a}nchen} shift from a weakly absorbing dielectric slab},}\ }\href {\doibase 10.1364/OL.30.002936} {\bibfield  {journal} {\bibinfo  {journal} {Opt. Lett.}\ }\textbf {\bibinfo {volume} {30}},\ \bibinfo {pages} {2936} (\bibinfo {year} {2005})}\BibitemShut {NoStop}%
\bibitem [{\citenamefont {Wang}\ \emph {et~al.}(2008{\natexlab{b}})\citenamefont {Wang}, \citenamefont {Ikram},\ and\ \citenamefont {Zubairy}}]{Wang_2008}%
  \BibitemOpen
  \bibfield  {author} {\bibinfo {author} {\bibfnamefont {Li-Gang}\ \bibnamefont {Wang}}, \bibinfo {author} {\bibfnamefont {Manzoor}\ \bibnamefont {Ikram}}, \ and\ \bibinfo {author} {\bibfnamefont {M.~Suhail}\ \bibnamefont {Zubairy}},\ }\bibfield  {title} {\enquote {\bibinfo {title} {{Control of the {Goos-H\"anchen} shift of a light beam via a coherent driving field}},}\ }\href {\doibase 10.1103/PhysRevA.77.023811} {\bibfield  {journal} {\bibinfo  {journal} {Phys. Rev. A}\ }\textbf {\bibinfo {volume} {77}},\ \bibinfo {pages} {023811} (\bibinfo {year} {2008}{\natexlab{b}})}\BibitemShut {NoStop}%
\bibitem [{\citenamefont {Ghaisuddin}\ \emph {et~al.}(2021)\citenamefont {Ghaisuddin}, \citenamefont {Abbas}, \citenamefont {Ali~Khan}, \citenamefont {Ali},\ and\ \citenamefont {Ziauddin}}]{Ghaisuddin-2021}%
  \BibitemOpen
  \bibfield  {author} {\bibinfo {author} {\bibnamefont {Ghaisuddin}}, \bibinfo {author} {\bibfnamefont {Muqaddar}\ \bibnamefont {Abbas}}, \bibinfo {author} {\bibfnamefont {Anwar}\ \bibnamefont {Ali~Khan}}, \bibinfo {author} {\bibfnamefont {Hazrat}\ \bibnamefont {Ali}}, \ and\ \bibinfo {author} {\bibnamefont {Ziauddin}},\ }\bibfield  {title} {\enquote {\bibinfo {title} {Enhancement of the {Goos-H\"anchen} shift in an optomechancal cavity via casimir force},}\ }\href {\doibase 10.1088/1402-4896/ac1dca} {\bibfield  {journal} {\bibinfo  {journal} {Phys. Scr.}\ }\textbf {\bibinfo {volume} {96}},\ \bibinfo {pages} {125104} (\bibinfo {year} {2021})}\BibitemShut {NoStop}%
\bibitem [{\citenamefont {Wild}\ and\ \citenamefont {Giles}(1982)}]{Wild-1982}%
  \BibitemOpen
  \bibfield  {author} {\bibinfo {author} {\bibfnamefont {Walter~J.}\ \bibnamefont {Wild}}\ and\ \bibinfo {author} {\bibfnamefont {C.~Lee}\ \bibnamefont {Giles}},\ }\bibfield  {title} {\enquote {\bibinfo {title} {{Goos-H\"anchen} shifts from absorbing media},}\ }\href {\doibase 10.1103/PhysRevA.25.2099} {\bibfield  {journal} {\bibinfo  {journal} {Phys. Rev. A}\ }\textbf {\bibinfo {volume} {25}},\ \bibinfo {pages} {2099} (\bibinfo {year} {1982})}\BibitemShut {NoStop}%
\end{thebibliography}%
	
\end{document}